\documentclass[review, colorinlistoftodos]{elsarticle}

\usepackage{lineno}
\usepackage[english]{babel}
\usepackage[unicode]{hyperref}
\usepackage{csquotes}
\usepackage{lscape}
\usepackage{physics}
\usepackage{amsmath}
\usepackage{tikz}
\usepackage{mathdots}
\usepackage{yhmath}
\usepackage{cancel}
\usepackage{color}
\usepackage{siunitx}
\usepackage{array}
\usepackage{multirow}
\usepackage{amssymb}
\usepackage{textcomp}
\usepackage{gensymb}
\usepackage{tabularx}
\usepackage{booktabs}
\usepackage{todonotes}
\usepackage[super]{nth}
\usepackage{ntheorem}
\usepackage{regexpatch}
\usepackage{xspace}
\usepackage{longtable}
\usepackage{enumitem}
\usepackage{graphicx}
\usepackage{ifthen}

\modulolinenumbers[5]
\newtheorem{hyp}{Hypothesis}

\usetikzlibrary{fadings}
\usetikzlibrary{patterns}
\usetikzlibrary{shadows.blur}
\usetikzlibrary{shapes}

\makeatletter
\regexpatchcmd\ps@pprintTitle
  {\cE\}\ \cE\}}{\cE\}\cE\}}
  {}{\FailedToPatch}
\makeatother

\makeatletter
\newcounter{subhyp} 
\let\savedc@hyp\c@hyp
\newenvironment{subhyp}
 {%
  \setcounter{subhyp}{0}%
  \stepcounter{hyp}%
  \edef\saved@hyp{\thehyp}
  \let\c@hyp\c@subhyp     
  \renewcommand{\thehyp}{\saved@hyp\alph{hyp}}%
 }
 {}
\newcommand{\normhyp}{%
  \let\c@hyp\savedc@hyp 
  \renewcommand\thehyp{\arabic{hyp}}%
} 
\makeatother

\newtheorem*{question}{RQ}

\addto\extrasenglish{%
}

\newcommand{\scrum}[1][0]{\ifthenelse{\equal{#1}{1}}{U}{u}se of Scrum\xspace}
\newcommand{\project}[1][0]{\ifthenelse{\equal{#1}{1}}{P}{p}roject success\xspace}
\newcommand{\wfh}[1][0]{\ifthenelse{\equal{#1}{1}}{H}{h}ome working environment\xspace}
\newcommand{\autonomy}[1][0]{\ifthenelse{\equal{#1}{1}}{N}{n}eed for autonomy\xspace}
\newcommand{\competence}[1][0]{\ifthenelse{\equal{#1}{1}}{N}{n}eed for competence\xspace}
\newcommand{\relatedness}[1][0]{\ifthenelse{\equal{#1}{1}}{N}{n}eed for relatedness\xspace}

\newcommand{\arrowr}{$\,\to\,$}
\newcommand{\ts}[1][0.3]{\hspace{#1cm}}

\begin{document}

\begin{frontmatter}

\title{The Impact of Working From Home on the Success of Scrum Projects: A Multi-Method Study}

 \author{Adrian-Alexandru Cucolaș}
 \address{Aalborg University, Denmark.
 \\
acucol19@student.aau.dk
 }
 
 \author{Daniel Russo}
 \address{Aalborg University, Denmark.
 \\
daniel.russo@cs.aau.dk
 }

\begin{abstract}
    The number of companies opting for remote working has been increasing over the years, and Agile methodologies, such as Scrum, were adapted to mitigate the challenges caused by the distributed teams. However, the COVID-19 pandemic imposed a fully working from home context, which has never existed before. This paper investigation a two-phased Multi-Method study. In the first phase, we uncover how working from home impacted Scrum practitioners through a qualitative survey. Then, in the second phase, we propose a theoretical model that we test and generalize using Partial Least Squares - Structural Equation Modeling (PLS-SEM) through a sample study of 200 software engineers who worked from home within Scrum projects. From assessing our model, we can conclude that all the latent variables are reliable and all the hypotheses are significant. Moreover, we performed an  Importance-Performance Map Analysis (IPMA), highlighting the benefits of the home working environment and the use of Scrum for project success. We emphasize the importance of supporting the three innate psychological needs of autonomy, competence, and relatedness in the home working environment. We conclude that the home working environment and the use of Scrum both contribute to project success, with Scrum acting as a mediator.
\end{abstract}


\end{frontmatter}

\section{Introduction} \label{sec:introduction}
With the recent COVID-19 pandemic, millions of people around the world were suddenly forced to either work from home or not at all, and companies were forced to allow their employees to work from home where possible in order to keep their businesses afloat \cite{impact_COVID}. Remote work has always been an alternative to being present in an office, especially within Software Development. A global survey shows that 56\% of companies allow remote working and 52\% of employees work from home at least once a week \cite{owllabs}. Now, with the rapid spread of the virus globally, countries imposed specific guidelines and restrictions regarding, for example, the number of people allowed in indoor spaces and their conduit, policies that, in this case, made working from home the only viable solution for many companies. However, despite apparent advantages such as reducing commute time and being able to spend more time with the family \cite{microsoft_challenges}, working from home is not an easy endeavor. It requires personal skills that allow the person to organize and carry his work without any form of direct supervision and that the job, i.e., the workplace, has a culture that supports and encourages remote work \cite{Baruch}.

Within Software Development, the rise of Agile methodologies in the latest decades has changed how people perceive and go about their work, as well as how companies encourage their employees to behave \cite{Agile_history}. According to the annual State of Agile report, Scrum was the most practiced Agile framework of all, and many companies already had some form of distributed Scrum within their organizations \cite{StateOfAgile}. Furthermore, the rate of project success using Scrum was reported to be 60\% higher than projects utilizing the traditional approach \cite{chaos_2018,scrum_evaluation_case_study}. However, the pandemic caused many companies to adapt to a fully remote work environment, putting Scrum practices under different strains. 

Studying the existing literature, we found that Scrum used for remote work has mainly been studied in the context of Distributed and Global Software Development, the focus being on how Scrum can be adapted to support a distributed working environment, commonly among co-located teams \cite{adapt_DNA}. However, the literature is scarce on how Scrum and the projects' success were affected by organizations switching to working fully from home to adhere to the new pandemic restrictions. That being said, our study aims to identify the impact that working from home had on software project success through Scrum practices. In our paper, the terms \textit{remote work} and \textit{working from home} are used interchangeably, as there is no other way of remote working than working from home during the pandemic. Hence, we want to clarify that throughout this study, our focus is on working from home rather than remote working.

Through a Multi-Method approach, we performed a qualitative survey with people working from home within software development Scrum projects to understand the phenomenon better. We proposed a theoretical model that we attempt to generalize through a sample study with 138 Scrum workers using Partial Least Squares Structural Equation Modeling.

In the remainder of this paper, \autoref{sec:related_work} discusses related work. First, the research design is presented in \autoref{sec:research_design}, illustrating the methodology this paper is based on. After that, \autoref{sec:phase_1} encapsulates the qualitative phase, referring to the exploratory study's data collection and analysis. Followed by the quantitative phase in \autoref{sec:phase_2}, where the model is constructed and evaluated. Next, the findings of our study are presented in \autoref{sec:discussion}, together with its limitations. Finally, \autoref{sec:conclusion} discusses future work as well as our conclusions.

\section{Related Work} \label{sec:related_work}
In the official Scrum Guide, Sutherland and Schwaber \cite{Scrum_guide} do not mention if Scrum can or cannot be used in the context of remote working. However, the literature on the topic argues that Agile practices, including Scrum, are meant for co-located working environments and that they present challenges in a distributed context, causing the necessity to adapt Scrum accordingly \cite{adopt_LR,scrum_dt_patterns}. Moreover, prior research has shown that ``only a small number of guidelines are systematically followed, and that some guidelines are rarely followed consistently'' \cite{use_scrum}, suggesting that each organization's Scrum implementation might differ from the guideline recommendations.

Prior research investigating Scrum concerning remote work targets distributed projects, also called Distributed Software Development and Global Software Development, where the focus is on how Scrum can be adopted and why it can be a good fit considering its benefits and challenges \cite{using_scrum_DAD,Agile_GSE}. The majority of papers discuss the challenges that distributed work creates in software engineering and how Scrum, with or without modifications, can address and mitigate these challenges \cite{ds_challenges}. The predominant theme is centered around co-located teams, mainly providing offshore software development services, which commonly implies working from different geographical locations within different time zones \cite{ds_outsource,why_scrum_works}. This requires Scrum to be adapted to support the coordination of the teams despite the challenges bridged by the geographical dispersion and the size of the projects \cite{adapt_GSD,mitigation_GSD,scale_scrum}. 

Another general topic in the literature is how Scrum can be scaled for Global Software Development since it generates new challenges such as communication and coordination overhead \cite{is_scrum_fit}. Some of the solutions found for mitigating the issues, as mentioned above, between multiple teams include: adding new events such as Scrum of Scrums for larger teams \cite{using_scrum_CS}, regularly visiting other sites for better process alignment \cite{ds_challenges}, using online tools for keeping the backlog and documentation up to date within teams \cite{anarchy} and conducting meetings online or via phone between remote teams \cite{scale_scrum}. 

Throughout the literature, researchers display ways of adapting and expanding Scrum practices to mitigate the challenges generated by distributed software development of geographically dispersed co-located teams \cite{using_scrum_SLR}. However, the perspective of everyone involved in a Scrum project working remotely, even within co-located teams, is missing from the literature. Some papers bring forward suggestions that could be adapted for this situation, such as virtual meeting rooms and fully remote Scrum events \cite{ds_challenges}. However, how the Scrum practices were impacted by the entire company working remotely remains to be thoroughly researched. For example, a recent action research study within a Brazilian start-up portrays how they managed to migrate to being fully remote to accommodate the COVID-19 restrictions \cite{COVID_startup}, yet a theoretical model remains to be developed and evaluated.

Moreover, most of the literature on distributed Scrum focuses either on successfully adapting or adopting the framework in a remote context, without accounting for the project's success. For example, Hassani-Alaoui et al. argue that ``if Scrum is easy to learn but difficult to master, scaled scrum is even more difficult to master, and conducting it improperly could have consequences for project success'' \cite{use_scrum}, suggesting that the adjustments made to accommodate the remote context may alter the project success.

Shared understanding, a challenge seen in the remote Scrum contexts caused by the communication impediments, has a significant effect on project success \cite{shared_understanding}. However, Hummel et al. ``did not find a significant moderation effect of team distribution on the effect of shared understanding on project success'' \cite{shared_understanding}, indicating that co-location is not necessary for effective communication as long as the proper tools are in place \cite{shared_understanding}. This finding suggests that besides the adaptions to Scrum, organizations must provide appropriate technological resources for remote contexts \cite{homework_COVID}.

Thus, our study differentiates from the existing literature by analyzing how Scrum practices are related to project success and what is the impact caused by the sudden change to remote working, which imposed everybody on working from home and eliminated the possibility of a co-located working environment.

\section{Research Design} \label{sec:research_design} 
\begin{figure}[h]
    \centering
    \includegraphics[width=\textwidth]{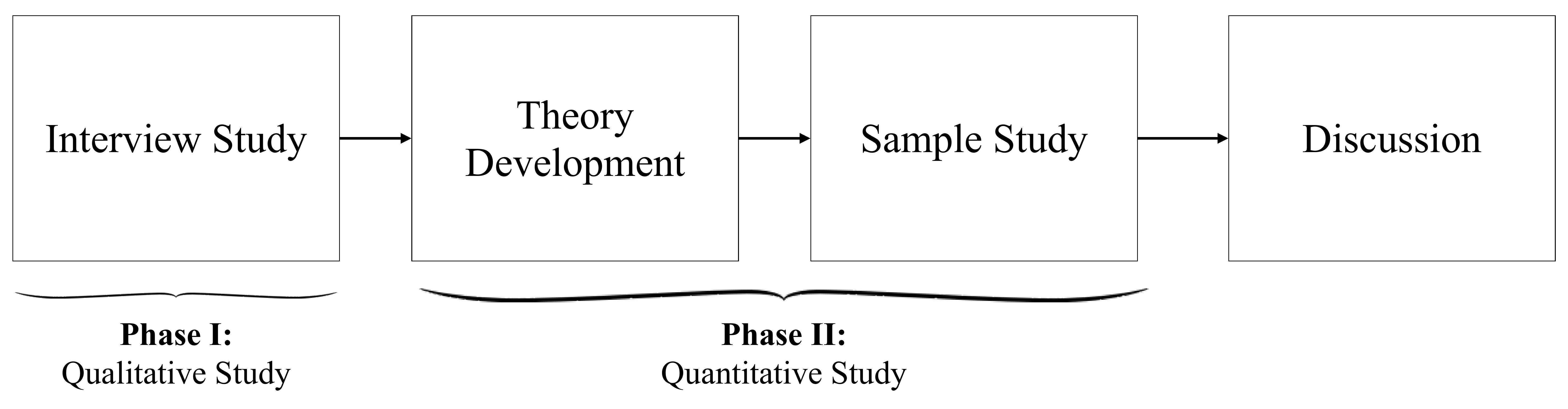}
    \caption{Design of the two-phased study}
    \label{fig:research_design}
\end{figure}

Our study aims to get a better understanding of the impact of working from home on Scrum and if it affects the project's success. For that, we employ a Multi-Method research approach, consisting of two phases as seen in \autoref{fig:research_design}. The benefit of using a Multi-Method approach is profiting from a research strategy's advantages while mitigating its weaknesses through the usage of another strategy \cite{mixed_method}.

For Phase I, we employed an exploratory qualitative survey approach as described in the ACM SIGSOFT empirical standards \cite{ACMStd}, consisting of semi-structured interviews with Scrum practitioners. As the prior knowledge about the topic under research is limited, the inductive technique is suitable \cite{content_analysis}. We aim to collect enough data in order to understand the phenomenon and quickly draw our findings directly from the responses \cite{confusion}. The semi-structured approach allows us to respond to and then ask further questions regarding what the interviewees answered, instead of completely following a predefined set of questions \cite{hear_data}. The responsive dynamic of semi-structured interviews aids in connecting with the interviewees and treating them as conversational partners rather than research objects \cite{hear_data}. To synthesize and analyze our findings, we use Gioia \cite{Gioia} as a data analysis technique. The qualitative phase is detailed in \autoref{sec:phase_1}.

Phase II complements and validates the qualitative study from the first phase by developing a theory and evaluating it through a quantitative sample study \cite{crowdsourcing_multi}. Within this phase, we employ a theory-building method to construct on the findings of the exploratory study as well as the literature, culminating in a theoretical model consisting of a set of hypotheses that are evaluated using Partial Least Squares - Structural Equation Modeling (PLS-SEM) \cite{pls_sem}. The hypotheses are evaluated on a sample study consisting of 138 Scrum workers. During this stage of research, the researcher is distanced from the observed constructs and statistically examines them \cite{Agile_success_model}. As the investigation's point of view is objective, and the hypotheses have been empirically validated, the study's findings are generalizable and independent of time, and context \cite{Agile_success_model,view_nowhere}. We discuss the quantitative phase in \autoref{sec:phase_2}.

We employ this research design to address the following research question:
\begin{question}
    Do Scrum practices affect software project success while working from home?
\end{question}

\section{Phase I: Qualitative Study} \label{sec:phase_1}
In the first phase, we followed ACM SIGSOFT empirical standards \cite{ACMStd} for conducting qualitative surveys and used Gioia \cite{Gioia} as the data analysis technique. This section presents the qualitative data collection and analysis.

\subsection{Qualitative Data Collection} \label{sub:qualitative_data_collection}
The pandemic context affected the data gathering process because the enforced lockdowns made it impossible to perform face-to-face interviews with participants even where geographically possible. For this reason, we had to adjust and create an online strategy for gathering interviewees and conducting the interviews.

The data is comprised of 12 interviews with people working remotely with Scrum within software development departments. To make it more enticing for people to participate in our study, we devised a small sign-up questionnaire (found in the Appendix, \autoref{fig:qualitative_survey}) that was meant to both filter out people who did not meet one of the requirements (i.e., having worked remotely within a Scrum project), and also let the participants choose their preferred interview time via a Calendly\footnote{https://www.calendly.com} link received upon completing the questionnaire and passing the criteria.

The questionnaire was posted on various social media websites, ranging from LinkedIn (and specialized Scrum-oriented LinkedIn Groups) to Facebook, and we also used our networks of family and friends to find interviewees. In addition, we contacted over 30 local Danish companies to participate, but unfortunately, none were interested.

The questionnaire was completed by 26 people, out of which four did not meet the requirements to participate. The 12 participants signed up through the link, and the rest of the ten qualified people received personal invitation e-mails but chose not to participate. An overview of the participants can be seen in \autoref{table:qualitative_participants}, and a snapshot of the questionnaire is available in \ref{appendix_a}. The interviews lasted between 20-35 minutes and were conducted and recorded via Zoom or Microsoft Teams, and then transcribed for analysis using Konch\footnote{https://www.konch.ai}. The transcripts were all further manually checked for possible errors created by the software, such as misinterpreted or missing words.

\begin{table}[h]
    \centering
    \small
    \begin{tabular}{@{\ts}l@{\ts[0.7]}l@{\ts[1]}c@{\ts[1.1]}l@{\ts[0.5]}}
    \toprule
    P\# & Scrum Role                 & Years of Scrum & Country   \\ 
        &                            & experience     &           \\ \midrule
    1   & Scrum Master               & 1-3            & India     \\
    2   & Developer                  & 1-3            & Romania   \\
    3   & Developer                  & 1-3            & Denmark   \\
    4   & Scrum Master               & 3-5            & Indonesia \\
    5   & Scrum Master/Product Owner & 5+             & USA       \\
    6   & Developer                  & 1-3            & Denmark   \\
    7   & Developer                  & 1-3            & Denmark   \\
    8   & Developer and Scrum Master & 3-5            & Denmark   \\
    9   & Developer                  & 5+             & Hungary   \\
    10  & Developer                  & \textless1     & Romania   \\
    11  & Developer                  & 1-3            & Spain     \\
    12  & Developer                  & 1-3            & Denmark   \\ \bottomrule
    \end{tabular}
    \caption{Participants distribution in the exploratory study}
    \label{table:qualitative_participants}
\end{table}

\newpage

The interview questions were designed to cover all aspects of Scrum - artifacts, events, and roles in comparison with working from home and on-site. Within the interviews, all the participants were asked about themselves, their workplaces and to describe their usual way of doing Scrum (most common before the lockdown), followed by a simulation of a current sprint, in the hopes of creating a space for them to think about and reflect upon their current Scrum process. The interview structure can be found in \ref{appendix_b}.

\subsection{Qualitative Data Analysis} \label{sub:qualitate_data_analysis}
To analyze the data, we used Gioia methodology \cite{Gioia} as a data analysis technique. However, given the qualitative exploratory direction of our research, throughout the process, we did not iterate between data collection and analysis or used theoretical sampling \cite{ACMStd}. Instead, we aimed to investigate how working from home affects Scrum projects inductively. For this reason, our goal in employing Gioia is to provide a systematic strategy that proved effective when performing our study and to assist readers in perceiving the rigor of our concept creation and progression to our conclusions.

We have begun the analysis by iteratively going through the transcripts and condensing information from the text in medium-sized codings while keeping informant terms (using participant's own words in the codings) to construct the \nth{1} order concepts \cite{Gioia}. The transcripts were coded with the research question in mind, focusing on what the participants voiced impacted their Scrum project the most while working from home. As such, each code summarizes the actual impact the interviewees observed, felt, or experienced remote work had regarding any role, event, or artifact of Scrum, as compared to how these practices were enacted before the COVID-19 pandemic. 

\begin{figure} 
    \centering
    \includegraphics[width=\textwidth,height=0.98\textheight,keepaspectratio]{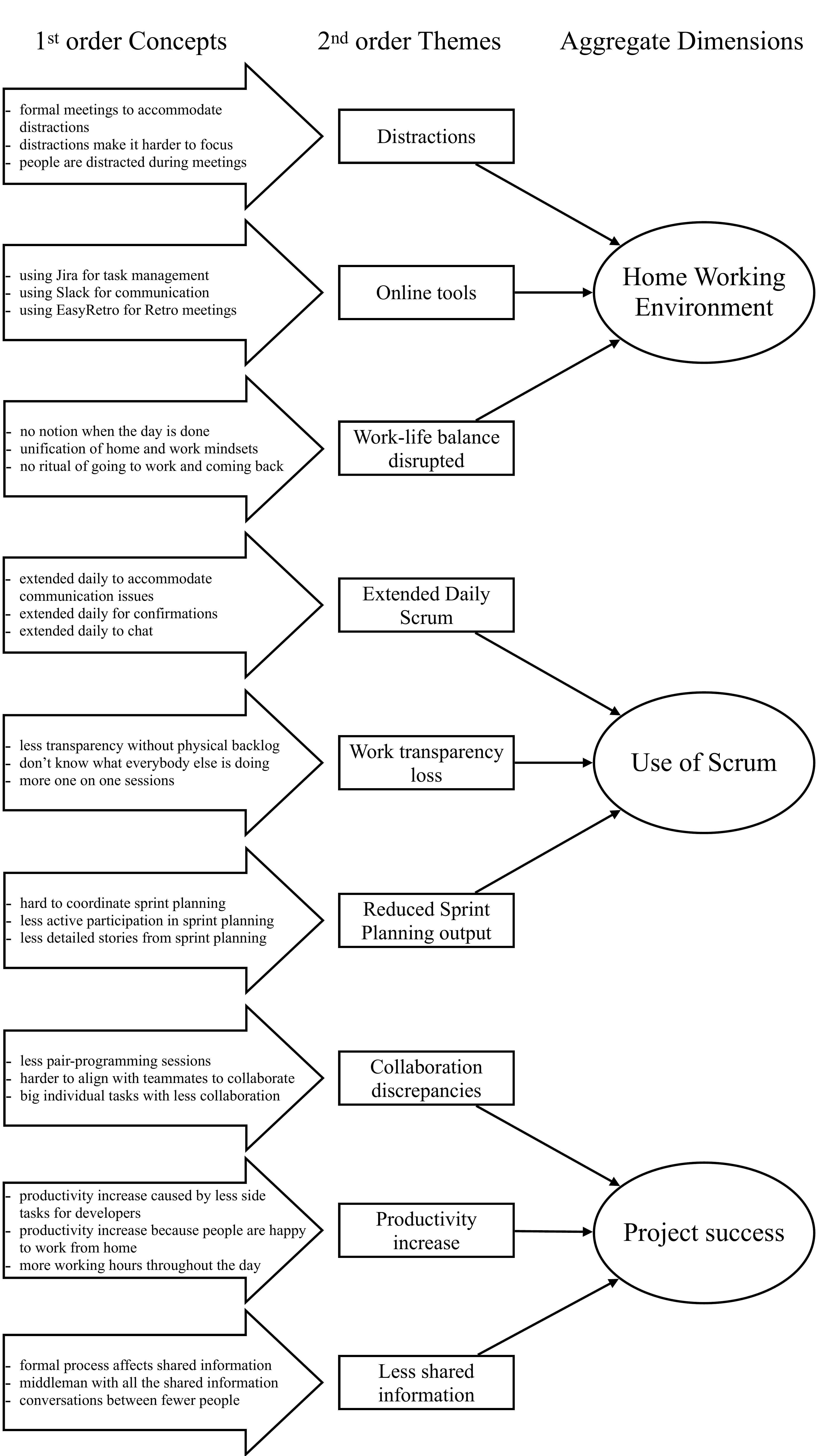}
    \caption{Data structure of \scrum, \project and \wfh}
    \label{fig:coding_scheme}
\end{figure}

\begin{figure} 
    \centering
    \includegraphics[width=\textwidth,height=0.98\textheight,keepaspectratio]{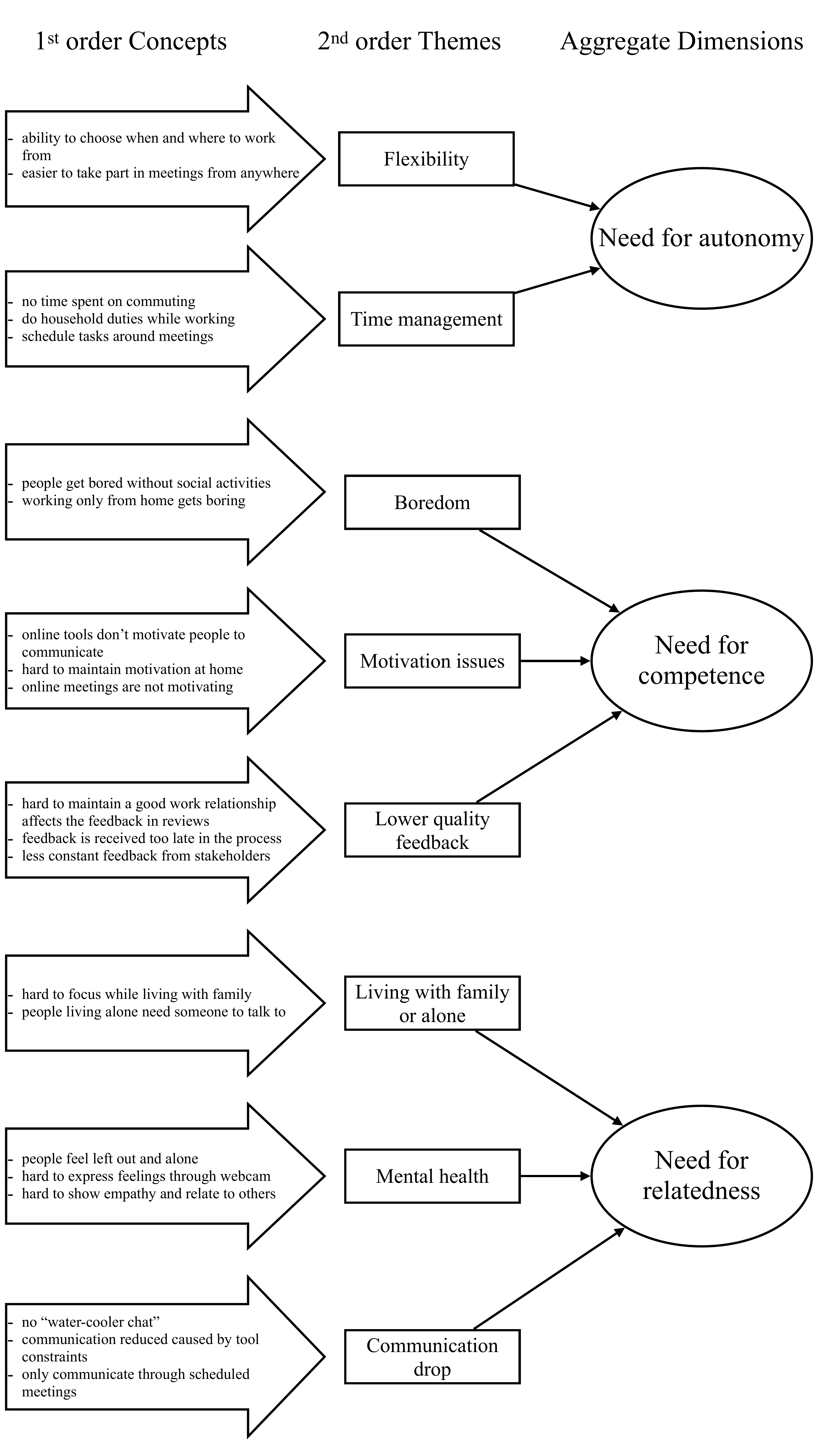}
    \caption{Data structure of the three innate psychological needs}
    \label{fig:coding_scheme_sdt}
\end{figure}

During the \nth{2}-order analysis, we begin to look for similarities and contrasts across the several concepts (akin to Strauss \& Corbin's \cite{gt_practice} concept of axial coding) in order to reduce their amount to a more manageable number \cite{Gioia}. The emerging themes will help us describe and explain the phenomena we observe \cite{Gioia}. Our analysis yielded a set of 17 \nth{2}-order themes seen as the underlying impacted areas, which were further distilled into six aggregate dimensions. The result is a data structure presented in \autoref{fig:coding_scheme} and \autoref{fig:coding_scheme_sdt}. The data structure not only helps us to construct our data into an understandable visual aid, but it also gives a graphic depiction of how we proceeded from raw data to \nth{1}-order concepts, \nth{2}-order themes and aggregate dimensions while completing the analyses, which represents an important component of proving rigor in qualitative research \cite{Gioia}.

The \scrum, \project and \wfh were each compiled as different dimensions (\autoref{fig:coding_scheme}) together with three additional ones (\autoref{fig:coding_scheme_sdt}) which reflect the impact of working from home on the individual through three innate psychological needs, further identified within the Self-Determination theory \cite{SDT}, i.e. \autonomy , \competence and \relatedness.

We uncovered that the impact of remote work was most visible on the Scrum Team, more specifically on the individuals comprising the Scrum Team. This finding allowed us to look at the events and artifacts through a different lens, making it clear that what the participants voiced about events or artifacts were simply manifestations of the affected individuals. In addition, we uncovered that the impact felt at the individual's level had implications in how the events were carried out, which also contributed to how the individuals were affected, creating a cycle that either positively or negatively affected Scrum as a whole. This concept in which working from home indirectly affected Scrum through its people was brought to life through the three psychological needs. The analysis made it more evident that working from home involves challenges for the individual, portrayed as three innate psychological needs: autonomy, competence, and relatedness. Russo et al. identified the exact three needs to be predictors of well-being and productivity among software developers working from home during the COVID-19 pandemic \cite{pandemic_predictors}. Moreover, together, the three psychological needs are "essential for facilitating optimal functioning of the natural propensities for growth and integration, as well as for constructive social development and personal well-being" \cite{SDT} as defined by Ryan et al. in their Self-Determination theory. Thus, the impact of working from home on the individual can be rightfully reflected through the three psychological needs.

Furthermore, the three dimensions related to the psychological needs indicate that working from home had a considerable effect on the individual, which is emulated into the use of Scrum and the project's success. Therefore, the remainder of this subsection details each dimension individually.

\subsubsection{\scrum[1]}
It cannot be generalized how the Scrum practices were impacted by working from home, as most companies have different ways of implementing and using Scrum within their organization \cite{real_world_scrum}. However, many interviewees voiced that their Daily Scrums were extended compared to the on-site version, which can be seen as a generalizable finding.
\begin{displayquote}[P4]
    Since the pandemic, the team decided to extend the time for daily Scrum. So after 15 minutes, they will close the daily Scrum and have extended 15 minutes to have a discussion as a confirmation session.
\end{displayquote}

While some companies only extended the scheduled duration of the Daily Scrum, other added additional events. The common reason being to enforce team communication and maintain the transparency of what everybody is doing, which is one of the three pillars of Scrum \cite{Scrum_guide}. Other reasons include allowing people from different teams to see each other using additional events such as Scrum of Scrums or just giving people the possibility of having an informal conversation before starting the workday. Although seen as something new as an effect of the sudden change to remote working by many people, additional and extended Daily Scrums are a common adaptation of using Scrum in a remote context, as noted in the literature \cite{using_scrum_SLR}. Furthermore, daily Scrums are an essential part of a healthy Scrum implementation; they "improve communications, eliminate other meetings, identify impediments to development for removal, highlight and promote quick decision-making, and improve the Development Team's level of knowledge" \cite{Scrum_guide}. 
\begin{displayquote}[P11]
    There are days in which I have been maybe working from home where out of the three people in the team, maybe with one or two of them, we didn't talk at all in the rest of the day after the daily.
\end{displayquote}

Given that the communication between the co-workers decreased throughout the day \cite{wfh_coworkers}, Daily Scrums became even more critical as some people would only get to communicate and share their progress through this event.

Regarding the other events, the interviewees perceived that their output was affected because of the decreased engagement and motivation of the participants. More particularly, the online experience of the Sprint Retrospective was not as pleasant as the on-site version for many interviewees. For this reason, Scrum Masters had to improvise and create new games to motivate the Development team and keep them engaged during the event. 
\begin{displayquote}[P12]
    Um, I don't like that (remote Retrospectives). We have done it a couple of times using online tools. But there's a component missing for me, like the human aspect, where you want to be honest and you got to express your feelings and all that.
\end{displayquote}

Moreover, the Sprint Planning meetings were also affected by the decreased engagement of the Development Team caused by the communication channel. Some interviewees voiced that their planning meetings were not as detailed as before because people would not pay attention. 
\begin{displayquote}[P12]
    You can hide easier and do something else when you're remote and when you're here (on-site office), you kind of have some sort of obligation to listen to the people in front of you. 
\end{displayquote}

The biggest challenge of the remote events perceived by the individuals in maintaining their focus on the topic and trying to engage in the conversation has its challenges while looking at or talking to a screen. Whereas in the on-site environment, the engagement is kept by a face-to-face conversation between the team, when working from home, it is very easy to get distracted and do something else on your machine while the other participants are talking.

The increased number of meetings had equally impacted the Product Owner and the Scrum Master.
\begin{sloppypar}
\begin{displayquote}[P4]
    What impacted me as Scrum Master. The level of my confidence dropped because I feel like I don't contribute anything to the company. And this happened because I need to have more meetings than before. For example, I cannot do something big because I have to accompany my product manager to have an ideation session. [...] So most of my days, I have a call rather than create something or observe the process.
\end{displayquote}
\end{sloppypar}

Both roles struggled to accommodate the new communication channels, which initially caused an increase in scheduled meetings. This, in turn, caused people to lose their confidence when unable to perform their role's duties due to the increased amount of meetings required to carry to support the business. Remote work has had a multi-sided impact on the Developers, consisting of many different underlying impacted areas. One element that was consistently missed by the participants is the physical interaction provided by sharing office spaces. Some argued that the interaction between the Developers changed because of working remotely: 
\begin{displayquote}[P3] 
    There was less interaction or less direct interaction, and you couldn't always see what everyone was doing, and they couldn't see what we were doing. Sometimes we had to wait until the demo or until the daily stand up next day to see progress.
\end{displayquote}

This has also resulted in a decrease in the quality of shared information and transparency among Developers, which has resulted in a decrease in the quality of the ambient knowledge intrinsic within teams.

Despite the product and sprint backlogs being two different artifacts within Scrum, no participant differentiated the two in our discussions. For most participants, the terms were interchangeable, and working with either backlog was also interchangeable due to the backlogs being in digital format in Jira or Trello. This is also why many participants felt no difference in interacting with the backlogs, as they were using online tools for managing their backlogs since before remote working, even though everybody used to have physical backlogs. However, some individuals miss the physical activity of interacting with the backlog, as having it in a digital format prevents them from having some sort of physical activity within a desk-job, as well as keeping them from quickly seeing a status of the Sprint, what each member is doing and who is available: 
\begin{displayquote}[P1] 
    I can see that there was a difference being done to physical activity like an on-site used to perform. I mean, sticking on the board, seeing what everyone was doing, like story points. [...] So that is the part which is missed.
\end{displayquote}
\begin{displayquote}[P4] 
    Before, in office, I created a physical board, right? So we can see other people, what they are doing—and moving the ticket, so they have some physical activity. But during this pandemic, they cannot see other people - what are they doing at this time? And when they need help, it's hard to get.
\end{displayquote}

Thus, the impact that working from home had on the use of Scrum is requiring the Scrum Team to consolidate and adapt the events, roles, and artifacts to mitigate the challenges caused by the remote context.  The adaptations regarding extended events, introducing new ones such as Scrum of Scrums, and employing online tools for task management are similar with the ones seen in the distributed Scrum literature \cite{ds_challenges}. This implies that Scrum can be applied to remote projects with no significant changes as long as people adhere to the framework's values and principles, but the official guidelines must be modified to better depict how Scrum can be used in a remote context.

\subsubsection{\autonomy[1]}
The \autonomy is an important aspect of having motivated employees. An autonomy-supportive context allows the individuals to feel competent, related, and autonomous, positively affecting their performance \cite{SDT}. In addition, working from home brings more flexibility over how and when the employees are completing their tasks, which is associated with more productivity and more engaged work to reward the employer for the flexible work arrangement \cite{wfh_coworkers}.
\begin{displayquote}[P12] 
    I am more relaxed when I get up in the morning, and this seems weird, but there is the option where you can say "I'm working from home," and then you can sleep one more hour because that's usually what you would spend to go to work. I think actually it gives you like a nice space buffer in your head saying like "OK, I'm going to work from home today because it seems like it's pretty windy". 
\end{displayquote}

Many people are forced to work outside of their prime hours by having to be present in the office at predefined hours, but working from home offers the freedom some seek in managing their own schedule and working when they feel more inclined to.
\begin{displayquote}[P6] 
    I see a huge time saving by not having to commute because it usually adds up to almost two hours every day. So, well, then I save two hours from my life every day.
\end{displayquote}
\begin{displayquote}[P8] 
    Something as simple as having 40 minutes of commute either way. And so those 40 minutes are spent working when I work from home.
\end{displayquote}

Many interviewees voiced that working from home saves precious time each day, which can be used for either working or doing something else not work-related. For those whose commuting time takes more than two hours, it eventually adds up to at least 10 hours spent on work-related activities in a day. Thus, working from home brings a great benefit by removing commuting altogether from the equation, allowing people to have more time for their personal life. Moreover, the flexibility and better time management resulted from working from home allows people to intertwine work and family life, giving them an advantage in meeting household needs \cite{domestic_space}.

On the other hand, working from home hinders the individual's interaction with his co-workers and the ability of the management to monitor the employees, creating more significant opportunities for work avoidance \cite{wfh_coworkers}. Thus, organizations should ensure contextual supports of autonomy for their employees and portrait working from home as an alternative to on-site development, rather than an escape from a controlled environment \cite{SDT}.

\subsubsection{\competence[1]}
 Together with optimal challenges, contexts related to social events, such as feedback and communication that aid the feeling of competence while doing a task, can intensify the motivation to complete the task \cite{SDT}. Working from home impacted the individual's motivation and feeling of achievement, challenging their perceived competence levels. As part of the Self-Determination theory, the \competence and \autonomy facilitate the vital expression of human growth tendency \cite{SDT}.
\begin{displayquote}[P4] 
    My confidence level dropped because I feel like I don't contribute something to the company, and my work is becoming smaller and smaller.
\end{displayquote}

The reduced amount of social interactions made people feel disconnected from the process and project objectives, leading them to detach even more from their work which they believed is insignificant. Other people perceived their tasks as boring when having to do them from home.
\begin{displayquote}[P2] 
    They were bored at home; they couldn't work from home anymore. They needed the office. 
\end{displayquote}

For some people, the seclusion makes them lose interest and motivation to complete their tasks and perceive their work as boring \cite{pandemic_predictors}. This is understandable for those working in functions or projects that require continuous communication, but it is also understandable for those who are lonely at home and miss the social context of the office, as lonely people report higher levels of boredom \cite{boredom_scale}. Either way, boredom is a result of their reduced motivation which affects their perceived competence.
\begin{displayquote}[P11] 
    I ask one person all the time and then imagine I make some code that's causing a problem. And it's not until the moment when this other person in the team, that knows about these corner cases, has to review it, and then he would give his feedback, and I would need to go back and redo and then show again. [...] It is more difficult to gather all the knowledge or all the feedback that you need until you know that the product is good.
\end{displayquote}

The async communication between fewer people impacts the individual's perceived competence, as feedback is received late in the process. The struggle to gather all of the necessary information and the lower quality feedback received throughout the completion of a task delays and obstructs the entire process of formulating and solving a task, perceived as difficult by the individual whose sense of competence is hampered.

\subsubsection{\relatedness[1]}
The \relatedness was challenged for some remote workers. Individuals are more likely to carry out activities that present interest for them or other individuals with whom they relate to \cite{SDT}. While working from home, most people perceived a great impact on the communication with their colleagues, which spawned different discrepancies in their usual way of working. For example, the ad-hoc conversations from the office became more formal. Given that just tapping on someone's shoulder and asking for help was no longer possible, meetings had to be scheduled and sustained through an online communication tool. Another observed impact was on individuals feeling left out or alone or needing someone to talk to. The Developers started initiating one-on-one meetings with their peers, but also with the Scrum Master or Product Owner in an attempt to maximize their well-being as individuals.
\begin{displayquote}[P4]
    We have more regular one-on-one sessions; even between the teams, they create a regular session with their peers. Before, when working from the office, they have rarely done that. I think they have one on one sessions because they need someone to talk to.
\end{displayquote}

The number of people an individual would have a conversation with during the day decreased to only a few close colleagues, or superiors \cite{wfh_coworkers}, forming more closed circles and increasing the bond between some people. However, affecting the overall transparency and the quality of the project's shared information while also discouraging individuals from starting conversations with other people outside their close circle. Moreover, some Developers' way of relating to the Scrum Master has also been changed compared to the on-site environment.
\begin{displayquote}[P2]
    There were times when he (Scrum Master) would have meetings and stuff like that, and he would just disappear from our place. This made us feel like he was way more superior or something like that, the fact that he could always leave the office. But during working from home, I didn't get that feeling at all because I wasn't seeing anyone around to notice if they're missing.
\end{displayquote}

As previously mentioned, while working from home, the individual interacts less with his co-workers \cite{wfh_coworkers}, making the perceived superiority of other people hinder as the individual is more focused on their work and also creating a sense of equality among the co-workers \cite{work_equality}. Another form of equality was brought up by the whole team communicating through telecommunication tools from their home, which made people relate more to their peers by having a small glimpse into everybody else's private life through seeing their home office set-up. Thus, the social aspects present in the office were diminished by people working from their homes. The findings are also supporting the literature, as the Self-Determination theory requires, besides the autonomy and competence needs, a third innate psychological \relatedness to produce motivation \cite{SDT}.

\subsubsection{\wfh[1]}
For many people, before COVID-19 caused everyone to work from home, they would not have an appropriate designated space for a home office set-up, making it harder to accommodate remote working at first. For this reason, people had to create a \wfh from where they can adequately do their job. Depending on their household situation, some people were forced to improvise an office and work from their sofas, beds or even kitchen tables \cite{pandemic_programming}. However, even though ergonomics are not significantly related to well-being or productivity \cite{pandemic_predictors}, improving comfort can help avoid future health issues. More importantly, to do their job, employees must have access to the proper virtual tooling.  
\begin{displayquote}[P6]
    Since our way of working was already entirely digitized, we had a digital scrum board, held meetings online, and so on, it wasn't big of a change.
\end{displayquote}

Luckily, most organizations were already using online tools for communicating and task management, which made the transition to working from home smoother \cite{virtual_workplace}. However, one big challenge of working from home is to avoid getting distracted and stay motivated \cite{pandemic_predictors}.
\begin{displayquote}[P5]
    If I'm on a call, we'll do our sprint planning, and just trying to get everybody's attention for more than an hour is hard. You have kids working at home, you know, on their schoolwork. You have dogs barking and all kinds of noises. Doorbell's ringing Amazon deliveries. So it's a lot more distracting, and then it's harder to get everybody back together if we have questions or need more information. 
\end{displayquote}

The participants voiced that it is difficult to avoid being distracted by what is going on around them unless they live alone or can appropriately seclude themselves from the rest of the people in their house. Everyone is focused on working at the office; however, this does not apply to the home working environment. The distractions can deteriorate the communication between the team members and have consequences on the overall project success. Another effect of the way people communicate when working from home is an increase in the number of formal meetings \cite{pandemic_predictors}, as time has to be better managed between people who only see each other through online tools. Meetings became more formal and scheduled to accommodate the asynchronous way of communicating online and avoid overlapping schedules. This gives people the impression that the number of meetings has increased because the previously ad hoc conversations have now become a time slot in their calendar.
\begin{displayquote}[P7]
    And then you also, of course, have the people that are late because they have children and their children have to go to school, or maybe they are arriving home, so they have to break it up in the middle of the daily stand up.
\end{displayquote}

The formal meetings are very defined on a topic and have to be scheduled to accommodate everyone's household duties schedule, particularly for the individuals living with their family. Another challenge brought up by working from home is the disruption of the work-life balance.
\begin{displayquote}[P10]
    I've had colleagues who said that they liked going to the office because it made it easier for them to have a schedule, and it brought more balance in their work and personal life in the sense that when they were working from home, they lost track of time. And most often, they worked too much, and they didn't realize that it was eight p.m. and they were still working.
\end{displayquote}

More particularly, without a clear working schedule, people got sucked in their work and stopped keeping track of how much they are working or used working as a coping mechanism for the ongoing pandemic \cite{pandemic_predictors}. This produced a spike in their perceived productivity as they were completing more tasks, but it affected their work-life balance. The ritual of having to commute to work and back was a clear delimiter for when to stop working for the day, yet it lacks in the \wfh.

\subsubsection{\project[1]}
The \project was also affected by the communication impact, as online communication is not as expressive as face-to-face, people had to optimize the way they share information between them. The communication became more formal and had to be more efficient and on point, but it affected some individuals' feelings. Especially making it harder for introverts to be included in social events as the former casual conversations and water cooler banter almost disappearing in the remote environment \cite{pandemic_predictors}.
\begin{sloppypar}
\begin{displayquote}[P7]
    The backend developers, we are cooperating less because someone is always working from home. We're not doing a lot of pair-programming, and we started taking on bigger tasks individually and working on those alone during the Sprint because then you don't really need to communicate a lot to each other.
\end{displayquote}
\end{sloppypar}

Since the communication between people was affected, their collaboration also presented some discrepancies. Popular on-site pair-programming sessions are less common remotely because people may better focus on their job and may no longer require assistance or find it more difficult to reach out to others for help. The collaboration discrepancies directly impact the project's success, as the shared feedback between co-workers decreased due to always having to communicate and collaborate through an online communication tool. When people are not in the same office, they tend to reach out less often for help as they never know what their colleagues are doing at that time \cite{wfh_coworkers}. Even if they do seek assistance, there is no way of predicting when the recipient will respond due to the async nature of the communication.
\begin{displayquote}[P5]
    I don't have to drive into the office anymore. I mean, I can turn my laptop on at seven a.m. and, you know, some days I finish off at four, some days I just work, maybe don't take a lunch. I mean, where am I going to go? So I think people are more productive because they can dedicate more time to working.
\end{displayquote}

People also observed a productivity improvement when working from home, which can be explained in various ways. A big struggle for most remote workers was to separate their work, and personal life \cite{pandemic_predictors}. Working from home created a shift in their work-life balance, and some people felt so connected to their work that they did not know when to stop working for the day. When working from the office, there is a clear distinction of when the workday ends, but working from home allows for more flexibility, which was hard to manage. Thus, one reason for the productivity spike is that people worked more hours \cite{wfh_hours} and the overall productivity of the project increased. However, this cannot guarantee if people were actually more productive or they had to work more to achieve the same results. Another reason is that working from home provides for better time management and helps employees to organize their work better. Because there were fewer work interruptions, the planned meetings, rather than ad hoc discussions, helped people focus on work for longer periods of time.
\begin{displayquote}[P1]
    The productivity even has increased because the colleagues are even happier while working from home and are given more hours.
\end{displayquote}
\begin{displayquote}[P9]
    I am more productive, that's for sure. These two things are connected because I'm more productive, which might boost myself because I've got this much amount of work done, and then I will work more because I'm boosted by this experience.
\end{displayquote}

Moreover, people felt better when they worked from home, and because well-being is related to productivity, the home environment had an influence on people's productivity, which is strongly tied to project success.
One obstacle that can endanger the project's success is the reduced amount of shared information resulting from the collaboration discrepancies.
\begin{displayquote}[P12]
    I think you're more prone to be the middleman where you kind of you write something to someone, and then you get a response and you write something to someone else and then you're the one sitting with all the information.
\end{displayquote}

Especially in a Scrum project, where transparency is one of the framework's pillars \cite{Scrum_guide}, the decreased amount of shared information can negatively affect its success. On the other hand, a streamlined communication method in which shared information is clear to all team members decreases the odds of errors and does not prolong the feedback process, boosting the quality of the development process and the project's likelihood of success.

\section{Phase II: Quantitative Study} \label{sec:phase_2}
In the second phase, we used existing Scrum literature and previous phase findings to develop a set of hypotheses that are integrated into a single theoretical model to improve our understanding of the relationship between working from home and the success of Scrum projects. The model is evaluated using Partial Least Squares – Structural Equation Modeling (PLS-SEM) following Russo \& Stol's guidelines \cite{pls_sem}. The goal is to use quantitative analysis to triangulate the findings of our qualitative phase. This section depicts the specification of the model and its assessment using a sample study and PLS-SEM.

\subsection{Theory Development} \label{sub:theory_dev}
The first goal of the quantitative phase is to develop the theoretical model, which will be evaluated with PLS-SEM.

\subsubsection{Partial Least Squares – Structural Equation Modeling} \label{subsub:pls}
 Partial Least Squares – Structural Equation Modeling (PLS-SEM) is a multivariate statistical analysis used to evaluate latent and unobserved variables, also called constructs, using multiple observable indicators or items \cite{pls_chin}. It is an emerging inquiry approach in empirical software engineering \cite{pls_sem,exploring_onboarding,Agile_success_model} and is best suited for exploratory theory development investigations \cite{soft_theory}.

 Every PLS-SEM model has two sub-models: a structural model and a measurement model \cite{pls_sem}. The structural model, as shown in \autoref{fig:model}, is made up of several constructs and their relationships define the research hypotheses. The latent variables might be exogenous (predecessor), endogenous (target construct) or mediators \cite{pls_sem,Agile_success_model}. The items are measured through the measurement model using data collected, most often, through a survey.

 We chose PLS-SEM for this study given its suitability for exploratory studies \cite{soft_theory} and to allow us to not only observe the direct relation between our constructs but also understand what causes the relation. Furthermore, we can evaluate the mediation effects between our latent variables, making PLS-SEM the best choice for our study.

\subsubsection{Specification of the Structural Model} \label{subsub:structural}
The first step in PLS-SEM is to specify the structural model by defining a set of hypotheses \cite{pls_sem}. The hypothesis formulation draws from both the literature and the qualitative study.

The working environment has an impact on the individual's motivation and productivity \cite{work_env_motivation,work_env_productivity}. Working from home has both positive and negative aspects. The positive ones include enabling employees to focus more on their tasks, as the contact with their co-workers is significantly reduced \cite{wfh_coworkers}. Research shows that while working from home, employees are interrupted less often \cite{telework_research}. Furthermore, teleworking gives employees more flexibility in terms of where they work from, how much work they do, and how they organize their time to complete their tasks even outside of the official office hours \cite{workplace_flexibility}. The results of the first phase show that employees felt more productive while working from home, which can be attributed to the autonomy that teleworking provides \cite{telework_effects}. While the findings from the exploratory study show that some employees did not feel motivated when working from home, this can be seen as a result of the ongoing pandemic, and the sudden change to enforced working from home \cite{pandemic_predictors}. Studies show that mixing online and offline working environments can help increase employees' motivation and engagement and that online communication tools are seen extremely well, especially by the Millennials who are now joining the workforce \cite{work_env_motivation}. As such, the hypotheses associating the three psychological needs with the \wfh are:
\begin{sloppypar}
\begin{subhyp}
    \begin{hyp}
        \autonomy[1] is positively associated with the \wfh[1].
    \end{hyp}
    \begin{hyp}
        \competence[1] is positively associated with the \wfh[1].
    \end{hyp}
    \begin{hyp}
        \relatedness[1] is positively associated with the \wfh[1].
    \end{hyp}
\end{subhyp}
\end{sloppypar}

The findings of the qualitative study show that working from home impacted the use of Scrum, particularly on some of its events. The literature on the topic argues that Agile practices, including Scrum, are meant for co-located working environments and that they present challenges in a distributed context, causing the necessity to adapt Scrum accordingly \cite{adopt_LR,scrum_dt_patterns}. As previously mentioned, the adaptations some Scrum events undergone while people had to work from home are similar to the adaptations that a distributed Scrum team has to enforce to support the communication between geographically dispersed co-workers \cite{adapt_GSD}. The working environment plays a big role in how people interact which each other, and working from home impacted the transparency between the people in the Scrum team, causing Daily Scrums to be extended or even requiring better documentation practices, effects also seen on distributed Scrum projects \cite{using_scrum_SLR}. The remote context has repercussions on the freedom and frequency of communication between team members, growing the chance of individualism, which threatens the \scrum \cite{scrum_factors}. Moreover, working from home brings difficulties related to teamwork, collaboration, and collective responsibility \cite{scrum_factors}. The \nth{2} hypothesis is:
\begin{hyp}
    \wfh[1] is negatively associated with the \scrum[1].
\end{hyp}

\project[1] is an important theme in project management literature, but it lacks a widely agreed definition of what success is \cite{project_success}. Traditional criteria such as time, budget, and quality, forming the "iron triangle," can be a good representation of the success of the management of the project but cannot necessarily reflect the success of the project itself \cite{project_success}. Shenhar et al. argue that between the three traditional dimensions of project efficiency (time, budget, and scope), the scope has the largest contribution in shaping the success of a project \cite{project_success_maping}. Thus, besides scope being an aspect of project efficiency, it also has an impact on the customer, and their satisfaction \cite{does_Agile_work}. In software development, the success metrics can also include stakeholder success \cite{stakeholder_project_success,stakeholder_multiple_model}, including satisfaction of the customer \cite{project_success_multidimensional}, team, meeting organizational goals \cite{project_success_framework} and strategic success/value \cite{project_success_multidimensional}. According to a global report \cite{chaos_2018}, it was found that the percentage of project success using the Scrum approach was 60\% greater than projects that use the traditional approach in 2018 \cite{scrum_evaluation_case_study}. While the above metrics are important to define the success of a project formally, the success factors of an Agile project might be nuanced differently as indicated by Russo \cite{Agile_success_model}. 

In an Agile project, such as Scrum, the success factors include Product Owner involvement, Scrum Master leadership, top management commitment, and Developers' skills \cite{Agile_success_model}. Ken Schwaber and Jeff Sutherland, the creators of Scrum, promote that "the essence of Scrum is a small team of people" \cite{Scrum_guide} and that Scrum is "a framework within which people can address complex adaptive problems, while productively and creatively delivering products of the highest possible value" \cite{Scrum_guide}. There is no silver bullet recipe for implementing Scrum, and each company can have varied ways of using Scrum \cite{using_scrum_SLR}. Moreover, even though there should be a Scrum Master to promote and support the good use of the framework within the project, the self-organizing Development Team (Developers) is responsible for managing their own work \cite{Scrum_guide}. Thus, the \scrum within a company is associated with the \project, stemming the \nth{3} hypothesis:
\begin{hyp}
    \scrum[1] contributes to \project[1].
\end{hyp}

The \wfh influences the way employees sustain their day-to-day work activities, which may, in turn, affect the \project. Building on the qualitative study, the communication, and collaboration between employees present challenges when people are working from home, findings also confirmed by the literature on remote Scrum projects \cite{scale_scrum,ds_challenges}. However, working from home is not always seen as a challenge for software engineers \cite{pandemic_predictors}, which was also the case for the people interviewed who indicated increased productivity while working from home. Moreover, people spend more time working instead of commuting when they are remote, which is directly affecting \project \cite{wfh_stick}. Thus, the \nth{4} hypothesis states:
\begin{hyp}
    \wfh[1] contributes to \project[1].
\end{hyp}

The literature suggests that using Scrum can help mitigate the challenges brought by remote contexts \cite{why_scrum_works}, indicating that the \scrum can be seen as a mediator between the \wfh and \project, rising the \nth{5} hypothesis:
\begin{hyp}
    \scrum[1] mediates the relationship between \wfh[1] and \project[1].
\end{hyp}

\begin{figure}[h]
    \centering
    \includegraphics[width=\textwidth]{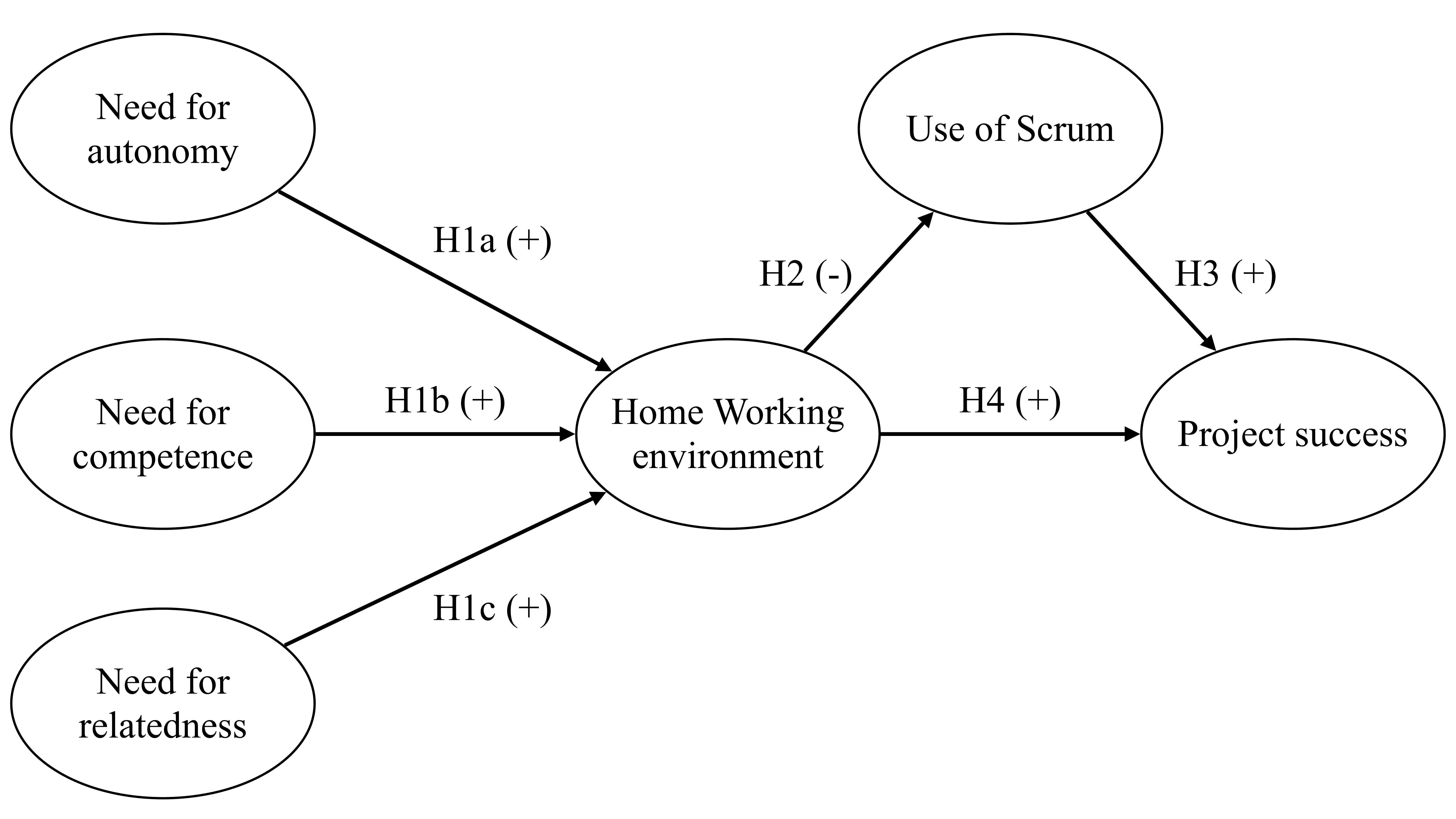}
    \caption{Proposed theoretical model}
    \label{fig:model}
\end{figure}

\autoref{fig:model} presents our theoretical model, where ovals represent the constructs and the hypotheses are represented by single-headed arrows.

\subsubsection{Specification of the Measurement Model} \label{subsub:measurement}
The second step is to specify the measurement model, that is, identify suitable measurement instruments for each non-observable variable \cite{pls_sem}. The model's constructs are defined below, indicating what each construct entails and what is its measurement instrument. The complete list of each construct's items together with the references used to frame the questions can be found in \autoref{table:items} in the Appendix.

All the responses were given on a 7-point scale ranging from 1 (strongly disagree) to 7 (strongly agree).

\paragraph{\textbf{\autonomy[1], \competence[1] and \relatedness[1]}}
The autonomy, competence and relatedness needs of the Self-Determination theory \cite{SDT} were measured using the 18-item balanced measure of psychological needs scale \cite{BMPN_scale}.

\paragraph{\textbf{\wfh[1]}}
The \wfh construct concerns distractions, noise, comfort, ergonomics, ambient, tooling, office set-up, and the social aspect of working from home. It is measured using nine items adapted from The Work Design Questionnaire (WDQ) \cite{wdq}, Danielsson et al. study of office environment satisfaction \cite{office_satisfaction} and Russo et al. study of predictors of well-being and productivity of software professionals \cite{pandemic_predictors} to reflect the home working environment.

\paragraph{\textbf{\scrum[1]}}
The \scrum concerns the implementation within an organization of Scrum's roles, events, and artifacts as defined in the official guide. It is measured using 15 items created from the Scrum guide \cite{Scrum_guide}. The responses reflect the extent to which the participants' Scrum implementation relates to the statements.

\paragraph{\textbf{\project[1]}}
The \project construct concerns the process, deliverable, and business success. It is measured using three items from Russo's Agile Success Model \cite{Agile_success_model}.

\subsection{Sample Study} \label{sub:sample_study}
After specifying our research model, we can concentrate on the empirical confirmation of our study hypotheses. Thus, we conducted a sample study, which is the best research strategy for generalizing a research model \cite{abc_se}. 

\subsubsection{Survey Data Collection}
We used G*Power \cite{g_power} to do a \textit{priori power test} to determine the minimal sample size. According to this analysis, the smallest size for six predictors with an effect size of 15\%, significance of 5\% and power of 80\% is 98 (\autoref{fig:g_power}).

Prolific\footnote{https://www.prolific.co/}, a data collection platform meant for academic purposes with over 146,000 active users, was used to gather the data using a cluster sampling approach \cite{behavioral_method,Agile_success_model}. Prolific offers various benefits over other data collecting platforms, such as email lists, replicability, data quality, and reliability \cite{prolific,platform_alternatives}, and is widely used in computer science as a data collecting platform \cite{Agile_success_model,pandemic_daily}. To reduce response bias, the survey was conducted using Qualtrics\footnote{https://www.qualtrics.com/} with randomized questions within their blocks \cite{behavioral_method,bias}. Following The Day Reconstruction Method (DRM) by Kahneman et al. \cite{DRM}, the questions within the survey were phrased such that people would reflect on their last project where they used Scrum while working from home.

To ensure the quality of the data, we added competence screening questions and random attention checks. The competence screening part included questions from Danilova et al. \cite{code_survey} and aimed to filter out people who do not meet the software engineering knowledge requirement. Attention checks were randomly inserted between questions to ensure the reliability of the answers. From a total of 200 completed responses, we filtered out 62 candidates who either failed the attention checks or the competence screening, summing up to 138 valid responses, which is above the minimum sample size.

\subsubsection{Sample Description} \label{subsub:sample_desc}
\ref{apx:sample_study} presents a detailed overview of our sample description. The gender distribution of the 138 respondents (\autoref{table:gender}) includes one non-binary respondent, while women make up 18.1\%, which is slightly higher than earlier sample studies that reported just 10\% of women participants \cite{open_source_impact,gender_online}. According to \autoref{table:country}, the great majority of the participants are from Western Europe.

In terms of Scrum, 86.2\% of the participants are part of the Development team (\autoref{table:scrum_role}), and most of them have more than one year experience of working in Scrum teams, $\approx$ 60\% even having at least three years experience (\autoref{table:scrum_exp}). This suggests that our participants are somewhat experienced with the Scrum practices. Moreover, \autoref{table:start_wfh} shows that most participants first started working from home when the COVID-19 pandemic restrictions were enforced. This benefits the study, allowing us to oversee the early impact that switching to working from home had on Scrum and the individuals of companies used to practice Scrum in an on-site environment.

\subsubsection{Assessment of the Measurement Model} \label{subsub:measurement_ass}
Following Russo \& Stol’s recommendations, the measurement model is evaluated before the structural one to ensure variables' reliability and robustness of the results \cite{pls_sem}. This section discusses internal consistency reliability, convergent validity, and discriminant validity.

\paragraph{\textbf{Internal Consistency Reliability}}
The tests ensure that the latent variables are measured in a \textit{reliable} and \textit{consistent} manner across their items. It aims to determine if the correlations between the items are high enough, indicating resemblance between the items of the same construct \cite{pls_sem}. We performed the Cronbach’s alpha \cite{cronbach_coefficient}, Composite Reliability \cite{composite_reliability} and Consistent Reliability Coefficient (rho\_A) \cite{consiste_reliability} tests. Values between 0.6 - 0.7 are acceptable in exploratory researches, while values between 0.7 - 0.9 are recommended, as they reflect satisfactory to good results \cite{pls_results_when}. On the other hand, values below 0.6 indicate a lack of internal consistency reliability, and values above 0.95 indicate that the items are similar, decreasing validity \cite{pls_sem}. As seen in \autoref{table:reliability_validity}, most constructs' tests fall within the recommended range of 0.7 - 0.9.
The only exceptions are \autonomy and \scrum with a Cronbach's Alpha and rho\_A above 0.6, which is acceptable given the exploratory nature of the research.

\begin{table}[h]
    \centering
    \small
    \begin{tabular}{@{\ts}l@{\ts[1]}c@{\ts[0.55]}c@{\ts[0.55]}c@{\ts[0.55]}c@{\ts[0.55]}}
    \toprule
    Construct   & Cronbach's & rho\_A & Composite   & AVE   \\ 
                & Alpha      &        & Reliability &       \\ \midrule
    \autonomy[1]    & 0.600      & 0.620  & 0.831       & 0.712 \\
    \competence[1]  & 0.808      & 0.828  & 0.885       & 0.720 \\
    \relatedness[1] & 0.709      & 0.741  & 0.836       & 0.632 \\
    \project[1]     & 0.725      & 0.731  & 0.845       & 0.645 \\
    \scrum[1]       & 0.611      & 0.612  & 0.793       & 0.561 \\
    \wfh[1]         & 0.789      & 0.803  & 0.862       & 0.610 \\ \bottomrule
    \end{tabular}

    \caption{Construct Reliability and Validity}
    \label{table:reliability_validity}
\end{table}

\paragraph{\textbf{Convergent Validity}}
This validity assessment measures the degree to which different items of the same construct correlate positively with one another \cite{pls_sem}. Given that all the model's latent variables are reflectively measured, the items should share a significant proportion of variance, meaning that they converge. To assess this assumption, two tests were performed: Average variance Extracted (AVE), which calculates the grand mean value of the squared loadings of the latent variable's indicators, and another one to check if the outer loadings of each construct share at least 50\% of their variance using the indicator's reliability \cite{pls_sem}. The AVE should score higher than 0.5 \cite{ave} and the outer loading of an item onto its construct should be higher than 0.7 \cite{indicator_realiability}. Several items did not share a proper amount of variance and were removed in a stepwise manner. Hair et al. suggest discarding all the items with an outer loading below 0.3 and consider discarding the ones with values between 0.4 - 0.7 if it leads to an increase in the AVE \cite{pls_primer}. Therefore, in turn, each lowest scoring indicator was discarded, then the model was re-run and verified that indeed the AVE value improved. We repeated this process until all the outer loadings passed the 0.7 threshold. \autoref{table:reliability_validity} shows that all AVE values are above the 0.5 threshold and \autoref{table:cross_loadings} displays the retained indicators' reliability scores, also passing the 0.7 threshold, through the cross-loadings. 

\begin{table}[h]
    \centering
    \small
    \resizebox{\textwidth}{!}{%
    \begin{tabular}{@{\ts}lcccccc@{\ts}}
    \toprule
    Item   & Need for       & Need for       & Need for       & Project        & Use of         & Home Working   \\ 
           & Autonomy       & Competence     & Relatedness    & Success        & Scrum          & Environment    \\ \midrule
    AN\_3  & \textbf{0.804} & 0.295          & 0.268          & 0.232          & 0.100          & 0.286          \\
    AN\_5  & \textbf{0.881} & 0.323          & 0.220          & 0.264          & 0.215          & 0.360          \\
    CN\_1  & 0.361          & \textbf{0.875} & 0.368          & 0.201          & 0.165          & 0.432          \\
    CN\_3  & 0.273          & \textbf{0.802} & 0.477          & 0.310          & 0.265          & 0.361          \\
    CN\_5  & 0.299          & \textbf{0.866} & 0.386          & 0.397          & 0.198          & 0.508          \\
    RN\_1  & 0.141          & 0.295          & \textbf{0.703} & 0.174          & 0.095          & 0.279          \\
    RN\_2  & 0.326          & 0.441          & \textbf{0.854} & 0.223          & 0.188          & 0.392          \\
    RN\_3  & 0.176          & 0.382          & \textbf{0.819} & 0.231          & 0.157          & 0.275          \\
    PS\_1  & 0.292          & 0.310          & 0.171          & \textbf{0.757} & 0.288          & 0.350          \\
    PS\_2  & 0.258          & 0.339          & 0.261          & \textbf{0.838} & 0.362          & 0.356          \\
    PS\_3  & 0.173          & 0.231          & 0.201          & \textbf{0.813} & 0.423          & 0.350          \\
    US\_12 & 0.029          & 0.213          & 0.163          & 0.370          & \textbf{0.766} & 0.174          \\
    US\_15 & 0.160          & 0.205          & 0.095          & 0.296          & \textbf{0.734} & 0.175          \\
    US\_5  & 0.244          & 0.130          & 0.160          & 0.339          & \textbf{0.746} & 0.260          \\
    HWE\_2 & 0.237          & 0.422          & 0.331          & 0.367          & 0.148          & \textbf{0.766} \\
    HWE\_4 & 0.373          & 0.431          & 0.353          & 0.422          & 0.269          & \textbf{0.836} \\
    HWE\_5 & 0.256          & 0.408          & 0.267          & 0.177          & 0.084          & \textbf{0.746} \\
    HWE\_6 & 0.321          & 0.368          & 0.306          & 0.351          & 0.309          & \textbf{0.772} \\ \bottomrule
    \end{tabular}%
    }
    \caption{Cross loadings of the retained items on the constructs}
    \label{table:cross_loadings}
\end{table}

\paragraph{\textbf{Discriminant Validity}}
This final validity test measures to which extent each construct uniquely captures different concepts about other constructs. We assessed this using the Heterotrait-Monotrait ratio of correlations (HTMT), which has been proven to outperform other tests, such as the Fornell-Larcker criterion \cite{HTMT}. It can be seen in \autoref{table:htmt_ratio} that all the cut-off values are below 0.9, which is the suggested threshold \cite{HTMT}, indicating that each construct is capturing a different phenomenon.

\begin{table}[]
    \centering
    \small
    \resizebox{\textwidth}{!}{%
    \begin{tabular}{@{\ts}lccccc@{\ts}}
    \toprule
    Construct                   & Need for & Need for   & Need for    & Project & Use of \\
                                & Autonomy & Competence & Relatedness & Success & Scrum   \\ \midrule
    \competence[1]  & 0.524    &            &             &         &                 \\
    \relatedness[1] & 0.419    & 0.627      &             &         &                 \\
    \project[1]     & 0.454    & 0.469      & 0.365       &         &                 \\
    \scrum[1]       & 0.306    & 0.354      & 0.274       & 0.664   &                 \\
    \wfh[1]         & 0.546    & 0.641      & 0.526       & 0.557   & 0.368           \\ \bottomrule
    \end{tabular}%
    }
    \caption{Heterotrait-Monotrait ratio of correlations (HTMT)}
    \label{table:htmt_ratio}
\end{table}

\subsubsection{Assessment of the Structural Model} \label{subsub:structural_ass}
We can conclude, from assessing the measurement model, that all our latent variables are reliable. The next step is to assess the structural model by discussing the predictive power and the significance of the relationships between the constructs \cite{pls_sem}. This evaluation is important for accepting or rejecting the proposed hypothesis \cite{pls_sem}.

\paragraph{\textbf{Collinearity}}
Our structural model consists of six constructs, out of which three are exogenous (\autonomy, \competence, and \relatedness). The correlation between the exogenous variables with the endogenous ones should be independent; for that, we assess their collinearity using the Variance Inflation Factor (VIF) \cite{vif}. A widely accepted cut-off value for VIF is 5, but Hair et al. suggests a more conservative cut-off value of 3 \cite{pls_primer}, as collinearity issues may appear between VIF values of 3 to 5. We report VIF values ranging from 1.245 (RN\_1) to 2.031 (CN\_1), which are below the more conservative cut-off value, indicating that our model does not present collinearity issues.

\paragraph{\textbf{Significance and relevance of path relations}} \label{p:significance}
The hypothesized relationships among the constructs are represented by path coefficients, which have standardized values ranging from -1 to 1 (for highly negative or positive correlations, respectively), while values close to 0 indicate a weak association \cite{pls_sem}. Because PLS-SEM does not make any distributional assumptions, parametric tests cannot be used to determine significance. For this reason, we used a 5,000-subsample bootstrapping approach with replacement, reported in \autoref{table:bootstrap}, where the bootstrapping coefficient, mean, standard deviation, T statistics, and \textit{p}-values are displayed for each of our six hypotheses.

Hypothesis 4 proposed that \scrum mediates the relationship between \wfh and \project. To evaluate mediating relationships, we must compare the mediators' suggested indirect paths to the direct paths \cite{mediate,mediation}. Variables can have no mediating impact (the indirect effect is insignificant), a partial mediating effect (if the direct effect is significant), or a full mediating impact (if the direct effect is insignificant) \cite{exploring_onboarding}. 

Both the direct (\textit{p}=0.000) and indirect (\textit{p}=0.006) paths are significant, indicating that \scrum has a partial mediating effect between \wfh and \project. Moreover, all the associations have \textit{p}-values less than 0.05, and the T statistic is more than 1.96 (for a significance of 5\%) \cite{pls_primer}. We conclude from this analysis that all our hypotheses are significant, thus supporting our research model.

\begin{table}[h]
    \centering
    \small
    \begin{tabular}{@{\ts}l@{\ts[0.4]}c@{\ts[0.7]}c@{\ts[0.7]}c@{\ts[0.7]}c@{\ts[0.7]}c@{\ts}}
    \toprule
    Hypothesis                & Coefficient & Mean & STDEV & T     & \textit{p} \\ \midrule
    \multicolumn{6}{c}{Direct Effects} \\
    H1a: NA \arrowr HWE & 0.204       & 0.207     & 0.074 & 2.769 & 0.006      \\
    H2a: NC \arrowr HWE & 0.360       & 0.360     & 0.081 & 4.448 & 0.000      \\
    H1c: NR \arrowr HWE & 0.176       & 0.189     & 0.077 & 2.280 & 0.023      \\
    H2:  HWE \arrowr US  & 0.274       & 0.286     & 0.087 & 3.159 & 0.002      \\
    H3:  US \arrowr PS   & 0.357       & 0.365     & 0.068 & 5.240 & 0.000      \\ 
    H4:  HWE \arrowr PS  & 0.340       & 0.343     & 0.069 & 4.942 & 0.000      \\ 
    \multicolumn{6}{c}{Indirect Effects} \\
    H5:  HWE \arrowr US \arrowr PS  & 0.098       & 0.104    & 0.036 & 2.743 & 0.006     \\ \bottomrule
    \end{tabular}
    \caption{Path coefficients, bootstrap mean, standard deviation, T statistics and \textit{p}-values (\mbox{NA = \autonomy[1]}, NC = \competence[1], NR = \relatedness[1], \mbox{HWE = \wfh[1]}, US = \scrum[1])}
    \label{table:bootstrap}
\end{table}

\paragraph{\textbf{Coefficients of determination and predictive relevance}}
In this step of the analysis, we are interested in the endogenous variables' predictive qualities, depicted in \autoref{table:coefficients_determination}. The $R^2$ value, or the coefficient of determination explained variance, is an essential metric in PLS-SEM since it determines the model's explanatory ability by determining how much of the variance is explained by each endogenous variable \cite{pls_sem}. Because the $R^2$ value is proportional to the model size, it is a good practice to account for the Adjusted $R^2$ criteria, which adjusts the $R^2$ value based on the model size \cite{r_adjusted}. The resultant number, which ranges from 0 to 1, indicates the amount of explanatory power. Threshold values are impossible to supply since they are dependent on the subject matter and the model's complexity \cite{pls_sem}. We also computed Stone-Geisser's $Q^2$ to see how predictive a given endogenous construct is \cite{cohen_statistical}. Finally, a blindfolding procedure was applied to compute $Q^2$ \cite{pls_primer}. The $Q^2$ must be greater than 0 to be relevant \cite{cohen_statistical}, which is the case for all endogenous constructs in our study. As a result, we draw the conclusion that our model is both predictively relevant and accurate.

\begin{table}[h]
    \centering
    \small
    \begin{tabular}{@{\ts}l@{\ts[2]}c@{\ts[1]}c@{\ts[1]}c@{\ts[0.6]}}
    \toprule
    Construct & $R^2$   & $R^2$   Adjusted & $Q^2$    \\ \midrule
    \wfh[1]       & 0.338 & 0.323         & 0.191 \\
    \project[1]   & 0.309 & 0.299         & 0.187 \\
    \scrum[1]     & 0.075 & 0.068         & 0.036 \\ \bottomrule
    \end{tabular}
    \caption{Coefficients of determination}
    \label{table:coefficients_determination}
\end{table}

\paragraph{\textbf{Predictive performance}}
To evaluate the predictive performance, we used the \textit{PLS predict} algorithm developed by Shmueli et al., which provides an assessment regarding a model’s predictive power by mimicking the evaluation of the predictive power for out-of-sample data \cite{pls_predict}. The process consists of dividing the data into $k$ equal-sized subsets, out of which $k-1$ are used to train the model, then its predictive power can be evaluated to assess whether the model can predict the $k^{th}$ subset \cite{pls_sem}. The goal of \textit{PLS predict} is to see if the model can outperform the most naïve linear regression benchmark (referred to as $Q^2$ predict). Thus a $Q^2$ predict value greater than 0 indicates that the PLS path model's prediction error is less than the naïve benchmark's prediction error \cite{pls_sem}. We divided our sample into 10 folds and used 10 repetitions to compute the \textit{PLS predict} statistics, following the guidelines by Shmueli et al. \cite{pls_predictive_model} to analyze the results. The predictive performance of the constructs is shown in \autoref{table:prediction_summary}. The positive values of $Q^2$ predict suggest a high predictive performance of the entire model. Moreover, all the indicators of the endogenous constructs were checked using the mean absolute error (MAE) as the prediction error, then double-checked using the root mean squared error (RMSE) against the naïve benchmark, and our conclusion does not change. Thus, the PLS prediction error for all items is lower than the naïve benchmark. As a result, we conclude that our model as a whole has a high predictive performance.

\begin{table}[h]
    \centering
    \small
    \begin{tabular}{@{\ts}l@{\ts[2]}c@{\ts[1]}c@{\ts[1]}c@{\ts[0.7]}}
    \toprule
    Construct & RMSE  & MAE   & $Q^2$ predict \\ \midrule
    \wfh[1]       & 0.860 & 0.664 & 0.293      \\
    \project[1]   & 0.960 & 0.708 & 0.131      \\
    \scrum[1]     & 0.991 & 0.765 & 0.053      \\ \bottomrule
    \end{tabular}
    \caption{Constructs prediction summary}
    \label{table:prediction_summary}
\end{table}

\paragraph{\textbf{Predictive Stability}}
The final assessment considers the model's predictive stability by analyzing the effect sizes ($f^2$). \autoref{table:effect_sizes} displays the effects of the different relations in the model. The effect size thresholds are 0.02, 0.15, and 0.35 for small, medium, and large effects, respectively \cite{cohen_statistical}. As such, the effect sizes in our model are all relevant, three of the relations having small effects and the other three medium effects.

\begin{table}[h]
    \centering
    \small
    \begin{tabular}{@{\ts}l@{\ts[1.5]}c@{\ts[1]}c@{\ts[1]}c@{\ts}}
    \toprule
    Construct   & Home Working & Project & Use of \\ 
                & Environment  & Success & Scrum  \\ \midrule
    \autonomy[1]    & 0.053        &         &        \\
    \competence[1]  & 0.140        &         &        \\
    \relatedness[1] & 0.036        &         &        \\
    \scrum[1]       &              & 0.171   &        \\ 
    \wfh[1]         &              & 0.155   & 0.081  \\ \bottomrule
    \end{tabular}
    \caption{Effect sizes ($f^2$)}
    \label{table:effect_sizes}
\end{table}

\begin{figure}[h]
    \centering
    \includegraphics[width=\textwidth]{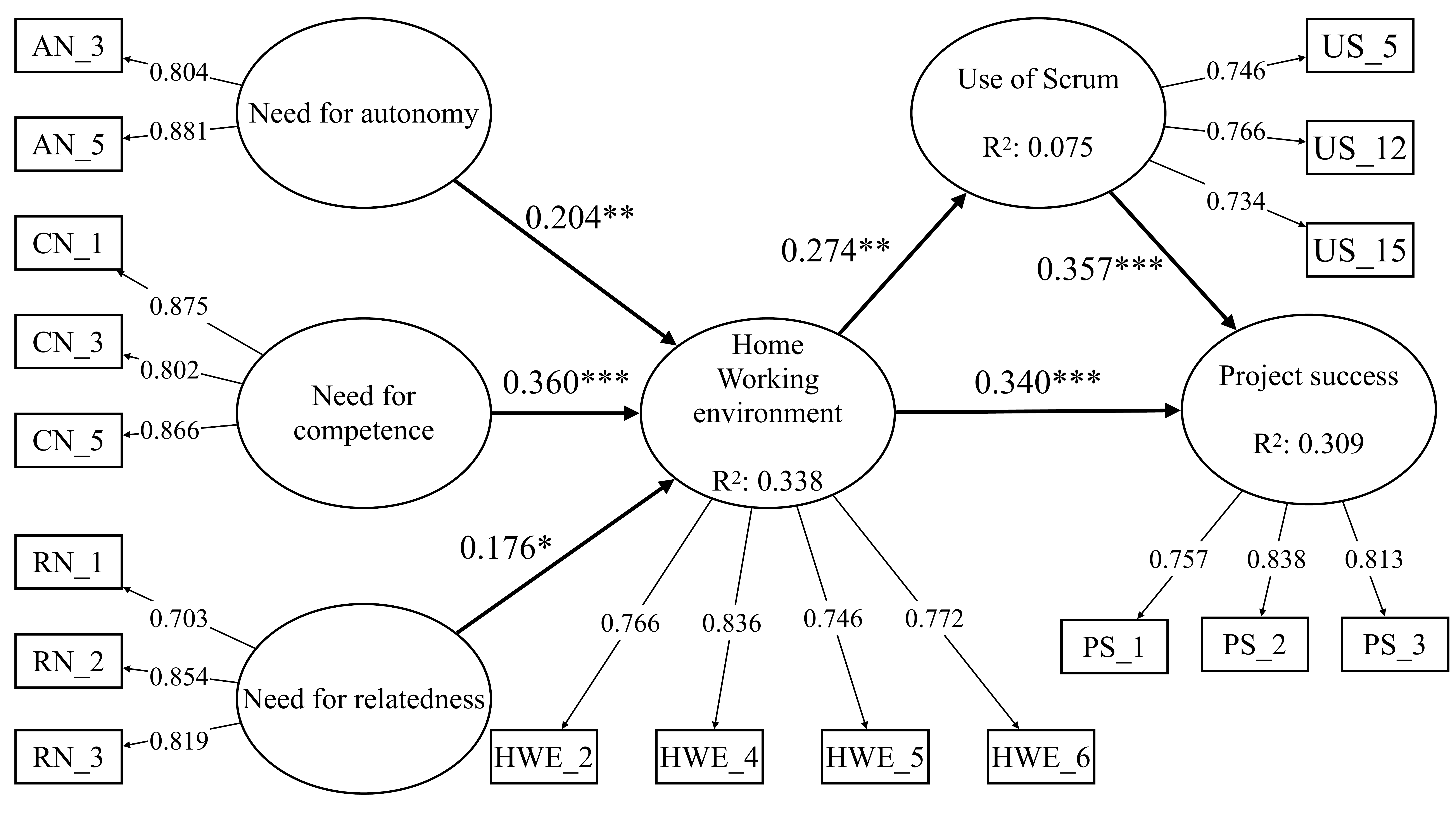}
    \caption{Measurement and Structural model with outer loadings, $R^2$, and path coefficients (*p\textless0.05, **p\textless0.01, ***p\textless0.001)}
    \label{fig:model_computed}
\end{figure}

Throughout this section, our research model was confirmed and validated through a PLS-SEM analysis. The computed model can be seen in \autoref{fig:model_computed}.

\newpage

\subsubsection{What is most important for success: an Importance-Performance Map Analysis} \label{subsub:IPMA}
This investigation focuses on the characteristics of \project in Scrum projects performed while people are working from home. ``The Importance-Performance Map Analysis (IPMA) is particularly useful for generating additional findings and conclusions by combining the analysis of the importance and performance dimensions in practical PLS-SEM applications'' \cite{ipma}. Furthermore, it enables determining the extent to which other latent variables improve \project. As a result, it gives managerial guidance in identifying the most important aspects of achieving \project in remote Scrum projects.

\autoref{table:performance} shows that all latent variables have good performance, ranging between $\approx$60\% and 75\%. This result is noteworthy, given that well-consolidated models, such as the technology acceptance model, exhibit construct performance ranging between 50\% and 70\% \cite{Agile_success_model,q_tam}.

\begin{table}[h]
    \centering
    \small
    \begin{tabular}{@{\ts}l@{\ts[2.5]}c@{\ts[1.9]}}
    \toprule
    Construct    & Construct Performances \\ \midrule
    \autonomy[1]    & 60.506                 \\
    \competence[1]  & 70.801                 \\
    \relatedness[1] & 59.577                 \\
    \project[1]     & 68.913                 \\
    \scrum[1]       & 65.994                 \\ 
    \wfh[1]         & 75.477                 \\ \bottomrule
    \end{tabular}
    \caption{Constructs Performance referred to \project}
    \label{table:performance}
\end{table}

The importance of individual constructs can be seen in \autoref{table:importance}. The values are comparable with those of other mature models (between 0.10 and 0.35) \cite{Agile_success_model,q_tam}. On a relative basis, we observe that \wfh is the most critical construct to \project (0.365), followed by \scrum (0.268).

\begin{table}[h]
    \centering
    \small
    \resizebox{\textwidth}{!}{%
    \begin{tabular}{@{\ts}lccc@{\ts}}
    \toprule
    Construct    & \wfh[1]   & \scrum[1] & \project[1] \\ \midrule
    \autonomy[1]    & 0.171 & 0.052 & 0.062   \\
    \competence[1]  & 0.388 & 0.118 & 0.142   \\
    \relatedness[1] & 0.127 & 0.039 & 0.046   \\
    \scrum[1]       & 0.000 & 0.000 & 0.268   \\ 
    \wfh[1]         & 0.000 & 0.305 & 0.365   \\ \bottomrule
    \end{tabular}%
    }
    \caption{Constructs Importance (Unstandardized Total Effects)}
    \label{table:importance}
\end{table}

\begin{figure*}[h]
    \centering
    \includegraphics[width=\textwidth]{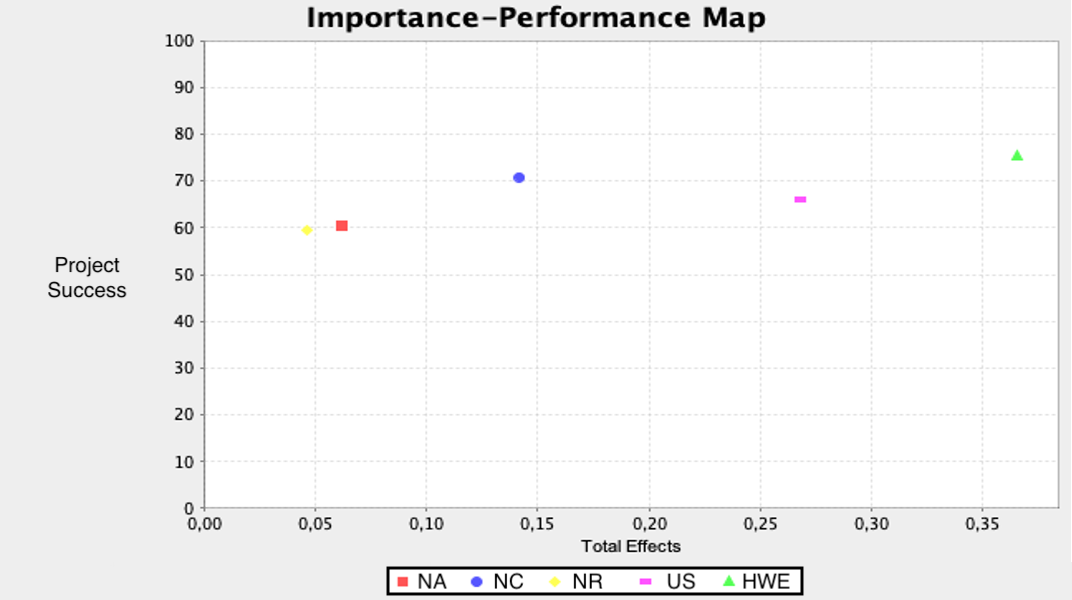}
    \caption{Importance-Performance Map Analysis of \project (\mbox{NA = \autonomy[1]}, NC = \competence[1], NR = \relatedness[1], \mbox{HWE = \wfh[1]}, US = \scrum[1])} 
    \label{fig:ipma}
\end{figure*}

The association between concept importance and performance is seen in \autoref{fig:ipma}. This figure's practical significance is that a one-unit point increase in home working environment’s performance increases the performance of \project by the value of home working environment’s total effect on \project, which is 0.365 (\textit{ceteris paribus}) \cite{ipma,Agile_success_model}. More precisely, if \wfh's performance would increase with one unit from 75.477 to 76.477, the performance of \project would increase by 0.365 points from 68.913 to 69.278. Thus, to increase \project, organizations should primarily focus on enhancing their employees' \wfh.

\section{Discussion} \label{sec:discussion}
The first phase (\autoref{sec:phase_1}), consisting of the qualitative research, helped us in better understanding how working from home affects the individuals within Scrum teams and how this translates into an impact on project success. We drew from these findings, together with the literature on software engineering, psychology, and project success, to develop our theoretical model in the second phase (\autoref{sec:phase_2}). Our model illustrates how these constructs relate to each other and helps us understand the role of Scrum within projects where people are working from home. Finally, we performed a sample study to generalize our findings and test our hypotheses. Following the most recent guidelines for software engineering research by Russo \& Stol \cite{pls_sem}, we analyzed the collected data through a PLS-SEM analysis as it is well-suited for evaluating a model's predictive abilities \cite{Agile_success_model,soft_theory}. All necessary statistical tests were passed by our model, demonstrating the significance and relevance of the results. \autoref{table:qualitative_implications} provides a summary of our findings and implications.

We found support for the hypotheses (\textit{H1a}, \textit{H1b}, \textit{H1c}) which proposed a positive relation between the three innate psychological needs and the \wfh. However, \autonomy and \relatedness had weak relationships with small effect sizes (both $\approx$0.05), while only \competence had a significant relationship with medium effect size (0.14). This may be explained by the fact that the \competence involves the \autonomy \cite{pandemic_predictors}. It is simpler to satisfy one's demand for competence while remaining relatively autonomous \cite{pandemic_predictors,SDT}. Moreover, the weak relationship of the \relatedness can be explained by the fact that, while working from home, people are less in contact with their peers \cite{wfh_coworkers}. Organizations must ensure that their employees' working environments, whether at home or on-site, support the three needs, according to the suggestions of the Self-Determination theory \cite{SDT}. Such an environment should encourage autonomy rather than control \cite{SDT}, offer employees the resources and training they need to feel competent about their work and make everyone feel heard and involved in the process.

We found a positive relationship between \wfh and \scrum, which does not support the proposed negative association of \textit{H2}. This finding confirms that Scrum can be adapted for distributed projects \cite{using_scrum_DAD} and it helps mitigate the challenges brought by the remote context \cite{ds_challenges}. Even though 72\% of our sample study participants first worked from home during the COVID-19 lockdown, most of them reported having the needed equipment and software to do their work, which suggests that the maturity of home working within the company is not related to the \scrum. Instead, it is crucial to have the right online tools in place, such that the employees have a smooth transition to remote working \cite{remote_Agile}. Moreover, the findings are consistent with the latest version of the Scrum Guide (2020) \cite{Scrum_guide}. The Scrum Guide became less prescriptive with the 2020 version, allowing more flexibility \cite{Agile_success_model}. For this reason, our paper contributes to the Scrum community by supporting the experience-based update of the framework \cite{Agile_success_model}. We recommend that businesses encourage their staff to experiment with various internet tools before deciding on the best ones. This is necessary to guarantee that the employees who work remotely are determined and find it simple to utilize these technologies to conduct their jobs.

As expected, we found a considerable positive contribution of the \scrum on \project. This finding is consistent with the literature, which shows that Scrum projects have a higher success rate than other traditional methods \cite{does_Agile_work,chaos_2018}. However, as also emphasized earlier, with the Scrum Guide 2020 \cite{Scrum_guide} being less prescriptive in terms of its practices and knowing that many organizations do not fully adhere to or adapt the practices \cite{use_scrum}, it is essential to follow the framework's values and pillars. Hassani-Alaoui et al. suggest that ``teams that frequently respected the pillars and values of scrum, even when modifications were made, appeared to achieve better outcomes'' \cite{use_scrum} in terms of \project.

\wfh[1] also has a considerable positive contribution on \project. This can be explained through the three innate psychological needs, which are also positively associated with \wfh. Ryan et al. state in their Self-Determination theory that ``employees' experiences of satisfaction of the needs for autonomy, competence, and relatedness in the workplace predicted their performance and well-being at work'' \cite{SDT}. Moreover, Russo et al. found that, among other predictors, \autonomy and \competence predicted the well-being of software developers during the pandemic \cite{pandemic_predictors}. Similarly, the IPMA suggests that the home working environment's performance related to \project is above 75\%, and its importance is the highest of the constructs (0.365). Thus, the \wfh offers a suitable setting in which developers may thrive, and enterprises should place a greater emphasis on improving their employees' working environments to achieve project success.

Finally, we found a partial mediating effect between \wfh and \project through the \scrum. Looking at IPMA, \wfh and \scrum had the highest importance levels related to \project with performance values of 75\% and 65\%, respectively. These findings support the literature's contention that Scrum aids in mitigating the difficulties associated with remote working \cite{using_scrum_SLR} while maintaining high rates of \project \cite{use_scrum}. Using a framework, such as Scrum, has the benefit of regulating the organization's process and, in the \wfh, it supports the employees to structure and plan their work around Scrum's events. This is important because employees could quickly lose focus and slack more while working from home without a straightforward process and defined goals. However, a controlled context hinders the competence and autonomy needs, affecting the performance of the employees \cite{SDT}. One way to cultivate a beneficial setting is to ensure that the current Scrum implementation within the projects adheres to the framework's core values (commitment, focus, openness, respect, and courage) \cite{Scrum_guide}, which support the three needs. As described in the Scrum Guide, ``the Scrum Team members are open about the work and the challenges; respect each other to be capable, independent people; and are respected as such by the people with whom they work and dare to do the right thing, to work on tough problems'' \cite{Scrum_guide}. In other words, the ideal Scrum Team is made up of autonomous, competent people who can take on challenging tasks and collaborate to reach their planned goal and attain \project. As a result of its flexibility in implementation and ability to accommodate changes, Scrum is well-suited to creating a beneficial \wfh that contributes to \project, as long as its values and pillars are followed.

\newpage

\begin{small}
    \begin{longtable}{@{\ts}p{2.1cm}p{4.3cm}p{4.3cm}@{\ts}}
        \toprule
        Hypothesis & Findings & Implications \\ 
        \midrule
        \textit{H1a}: \autonomy[1] \arrowr \wfh[1] & Supported. With a low path coefficient (0.2), \autonomy is positively associated with \wfh, even though the effect size is small (0.05). & The autonomy of working from home can result in more productivity and more engaged work. Organizations should ensure an autonomous context, rather than a controlling one. \\
        \addlinespace
        \textit{H1b}: \mbox{Need for} \mbox{competence} \arrowr \mbox{Home working} environment & Supported. This relationship is stronger than H1a, with a considerable path coefficient (0.36) and a medium effect size (0.14).  & While working from home, people took on more individual tasks which challenged their competence. Organizations should establish a streamlined feedback process with high degree of shared information.  \\
        \addlinespace
        \textit{H1c}: \relatedness[1] \arrowr \wfh[1] & Supported. Weak relationship between the \relatedness and \wfh with a low path coefficient (0.17) and small effect size (0.03). &  People did not relate as much with their colleagues when working remotely. Organizations should devise more social events to keep people connected. \\
        \addlinespace
        \mbox{\textit{H2}: Home} working \mbox{environment} \arrowr Use \mbox{of Scrum}  & Not supported. \wfh[1] is not negatively associated with \scrum. However, there is a positive association through a moderate path coefficient (0.27)  with small effect size (0.08). & Scrum can successfully be adapted for remote working. Organizations have to accommodate the remote challenges by adopting and training their employees to use online communication and task management tools. \\
        \addlinespace
        \mbox{\textit{H3}: Use of} \mbox{Scrum \arrowr} \mbox{\project[1]}  & Supported. \scrum[1] has a statistically significant and considerable positive association with \project (path coefficient of 0.35 and moderate effect size of 0.17). & Scrum comes in many flavors depending how it is used to achieve project success. However, organizations must ensure that their Scrum implementation respects the pillars and values of Scrum.   \\
        \addlinespace
        \mbox{\textit{H4}: Home} working \mbox{environment} \arrowr \project[1]  & Supported. \wfh[1] also has a statistically significant and considerable positive association with \project (path coefficient of 0.34 and moderate effect size of 0.15). & The workers' experience in achieving the needs for autonomy, competence and relatedness predicts their work performance and well-being. Organizations can do better by enhancing their employees' \wfh. \\
        \addlinespace  
        \mbox{\textit{H5}: Home} working \mbox{environment} \arrowr \mbox{Use of} Scrum \arrowr \project[1]  & Supported. \scrum[1] has a partial mediating effect between \wfh and \project (path coefficient 0.09, \textit{p} = 0.006). & Using Scrum helps mitigate the remote working challenges which could hinder project success. Organizations should consider using Scrum within their remote teams to take advantage of the framework's benefits which are also reflected on \project. \\
        \bottomrule
        \caption{Summary of findings and implications}
        \label{table:qualitative_implications}
    \end{longtable}
\end{small}

\subsection{Threats to Validity} \label{sub:limitations}
Both qualitative and quantitative validity paradigms are used to discuss the limitations of the study. First, the qualitative dimension is analyzed by considering the credibility, transferability, dependability, and confirmability following the Criteria for assessing the trustworthiness of naturalistic inquiries \cite{trustworthiness_criteria}. Second, for the quantitive dimension, its evaluation investigated the statistical conclusion, internal, construct, and external validity as suggested by Wohlin et al. \cite{experimentation}.

\paragraph{\textbf{Credibility}} \label{p:credibility}
Our findings are based on a qualitative survey approach as described in the ACM SIGSOFT empirical standards \cite{ACMStd}, consisting of 12 interviews with individuals from three continents and seven countries, with various Scrum experience levels, working in all existing Scrum roles. The participants were screened before the survey, and those who did not meet the requirements of working from home within a Scrum project were filtered out. 
Additionally, the sample size is in line with other studies applying a Mixed Methods approach \cite{Agile_success_model}.

\paragraph{\textbf{Transferability}} \label{p:transferability}
According to Gioia et al., the findings of a study based on this methodology are transferable as long as the study ``generates concepts or principles with obvious relevance to some other domain'' \cite{Gioia}. Although we only used Gioia as a data analysis technique, based on this notion, the findings within this paper, while limited in its sample size, can be considered transferable, as the underlying impacted areas discovered throughout the study are human-centric, therefore branching out into other sciences such as social sciences, psychology or anthropology. As such, the findings of this study can be used as a basis for further investigation of similar matters, therefore enriching the academic understanding of how working from home impacts Scrum knowledge workers. However, this study focuses on Scrum practices; thus, we cannot make any assumptions for projects using other Agile frameworks or a combination.

\paragraph{\textbf{Dependability}} \label{p:dependability}
The research process is consistent, and the data is stable. An external auditor can examine the process following the ``audit trail'' \cite{trustworthiness_criteria} consisting of documentation gathered by the authors throughout the process, along with the actual data.

\paragraph{\textbf{Confirmability}} \label{p:confirmability}
We guided ourselves by the ACM SIGSOFT empirical standards \cite{ACMStd}, which provided a broader understanding of what a qualitative study entails, allowing us to be less constrained in our analysis process. As such, our study provides a transparent explanation of our analysis process, with a clear chain of evidence from the participant (data) to the proposed concepts \cite{ACMStd}.

\paragraph{\textbf{Internal}} \label{p:internal}
We used a cluster-randomized probability sampling strategy to validate our model \cite{behavioral_method}. Thus, we only used a cluster, namely the Prolific community, instead of the entire world population, which would not be feasible. We recognize that cluster sampling is less precise compared to random sampling, but it is far more cost-effective than other sampling strategies \cite{Agile_success_model}. Furthermore, we added screening questions and random attention checks to filter out inadequate responses to enhance the data quality. With this strategy, we filtered out 62 candidates out of 200 who did not meet our requirements, concluding to 138 valid responses. However, as most of our informants are from Europe, we recognize that our sample is not representative of the software engineering population employing Scrum.

\paragraph{\textbf{External}} \label{p:external}
Because sample studies are ideally suited for theory generalization, the fundamental goal of the PLS-SEM study has been the generalization of our findings \cite{abc_se}. We received 138 valid responses, which was more than acceptable given the results of the \textit{priori power test} we conducted using G*Power \cite{g_power} before the data collection (suggesting a sample size of 98 participants). However, our model only accounts for Scrum projects, thus cannot be generalizable for projects using other Agile frameworks.

\paragraph{\textbf{Construct}} \label{p:construct}
Using a single-informant approach, the latent variables were measured to reflect a Scrum worker's perspective \cite{Agile_success_model}. We solely employed self-reported metrics, asking our informants to rate their level of agreement with literature-derived indicators. To compensate for questions not being correctly answered, we added random attention checks to filter out dishonest candidates. Furthermore, we adapted existing measurement instruments from the literature \cite{Agile_success_model} and also created the \scrum measurement instrument from the official Scrum guide \cite{Scrum_guide}, as no validated scale was identified. For this reason, 26 out of 45 indicators were discarded from the model due to their poor loadings.

\paragraph{\textbf{Statistical conclusion}} \label{p:stat_conclusion}
The survey findings were computed using the Partial Least Squares – Structural Equation Modeling method with the well-known statistical tool SmartPLS (v. 3.3.3), which has been used in over 1,000 peer-reviewed studies \cite{Agile_success_model,smartpls}. Furthermore, all statistical techniques and tests utilized in the PLS-SEM analysis are in accordance with the most recent recommendations in our area \cite{Agile_success_model,pls_sem}.

\section{Conclusion} \label{sec:conclusion}
Using Scrum in globally distributed projects can help mitigate the challenges brought by the remote context. The literature already focused on the best adoption and adaptation strategies towards distributed Scrum. However, the pandemic generated a context where the entire organization is working remotely, which has yet to be researched. Therefore, our study aimed to uncover the impact of working from home on the success of Scrum projects.

We employed a two-phased Multi-Method approach to better understand the impact through an inductive qualitative survey with Scrum practitioners who worked remotely. Then, we formulated a theoretical model drawing from our findings and the literature. We evaluated the model using Partial Least Squares - Structural Equation Modeling through a sample study of 138 participants to generalize our findings. To better understand what drives \project, we performed an Importance-Performance Map Analysis (IPMA), highlighting the benefits of the \wfh and \scrum for \project. We found that the \wfh nurtures the three innate psychological needs of the employees' autonomy, competence, and relatedness, making them perform better while working from home. Moreover, the \scrum mediates the \wfh and \project, suggesting that remote projects using Scrum have higher chances of attaining \project.

This study contributes to the limited literature within software engineering on working from home and its relation to Scrum and \project. While our investigation was purely exploratory, future research can build on our findings to create more fine-grained theories to fully understand the impact of working from home on the current software engineering community. Moreover, it would be insightful to expand our model on the Agile methodology to capture a broader understanding of the current phase of the industry. Finally, future studies might build on our findings by assessing the relationship between Scrum's values and principles and the three innate psychological needs to provide further suggestions on creating a beneficial working environment for employees, which also reflects in \project.

\section*{Supplementary Materials} 
Raw data and computation tables are openly available under a CC BY 4.0 license on Zenodo, DOI: \url{https://doi.org/10.5281/zenodo.4924061}.

\section*{Acknowledgments}
This work was supported by the Carlsberg Foundation under grant agreement number CF20-0322 (PanTra --- Pandemic Transformation).
 The authors would also like to thank Ioana-Mădălina Gănescu for helping gather and analyzing some interviews.

\newpage

\bibliographystyle{elsarticle-num}
\bibliography{mybibfile}

\newpage
\appendix

\section{Qualitative Survey}\label{appendix_a}
\begin{figure*}[!ht]
\centering
\includegraphics[width=\textwidth]{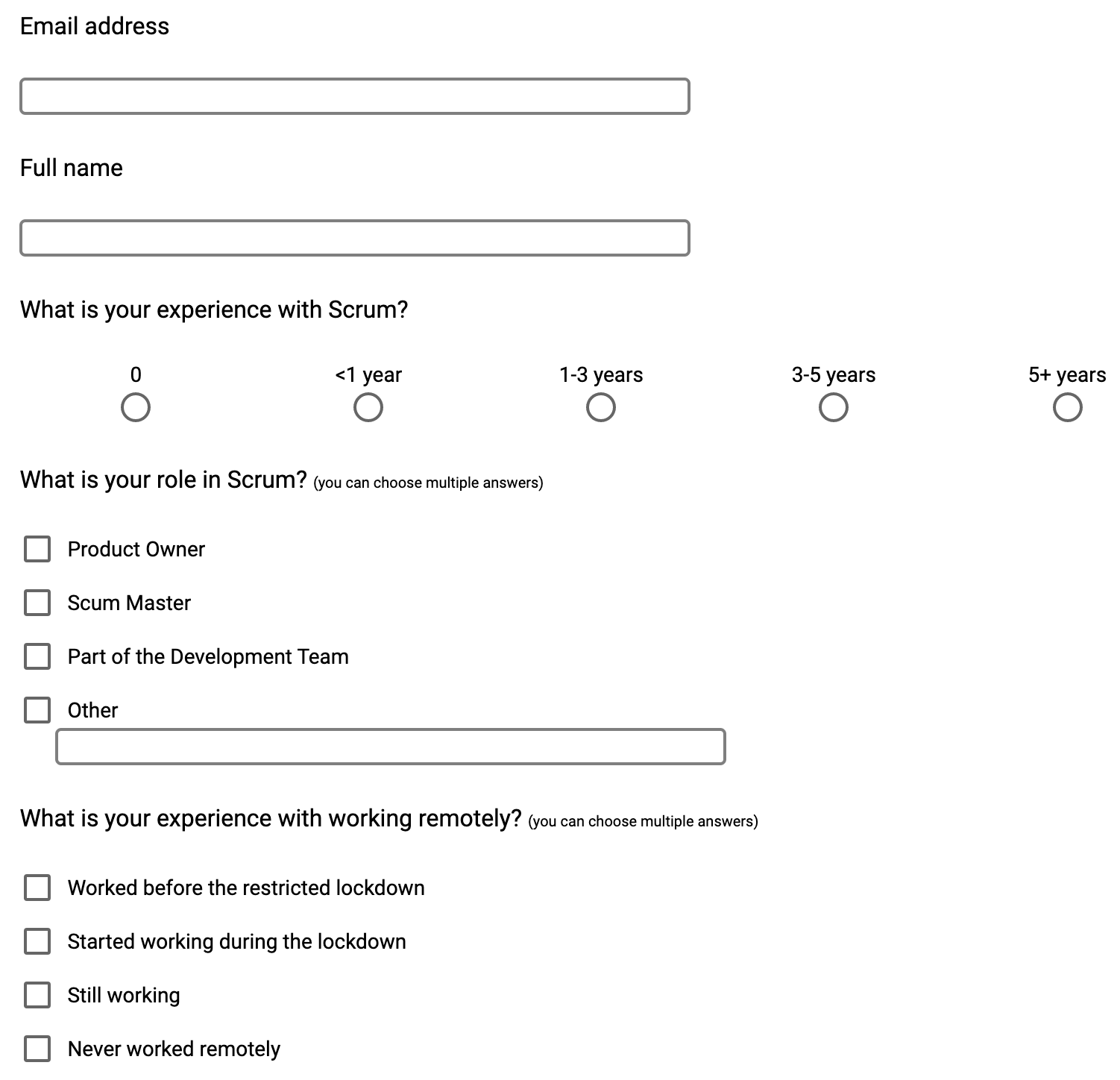}
\caption{Qualitative Survey to filter out participants who do not meet the research requirements} 
\label{fig:qualitative_survey}
\end{figure*}

\section{Interview Structure}\label{appendix_b}
\begin{table}[h]
    \centering
    \small
    \begin{tabular}{@{\ts}l@{\ts[2.5]}l@{\ts[2.5]}l@{\ts[1]}}
    \toprule
    Scrum Practice           & On-site & Remote \\ \midrule
    Daily Scrum              &         &        \\
    Sprint Review            &         &        \\
    Sprint Retrospective     &         &        \\
    Backlog Refinement       &         &        \\
    Sprint                   &         &        \\
    Task Breakdown           &         &        \\
    Requirements Elicitation &         &        \\
    Estimation               &         &        \\
    Task Assignment          &         &        \\
    Product Owner            &         &        \\
    Scrum Master             &         &        \\
    Developers               &         &        \\
    Backlogs                 &         &        \\
    Others                   &         &        \\ \bottomrule
    \end{tabular}
    \caption{Interview structure}
    \label{table:interview}
\end{table}
 
 For each Scrum Event in \autoref{table:interview}, three questions were always asked for both On-site and Remote contexts:
 \begin{enumerate}
     \item What is the process?
     \item How long does it take?
     \item Who participates?
     \item What is the output? Is there a difference between the contexts? 
     \item How do you feel about doing this event in this context?
 \end{enumerate}

\section{Items description}\label{apx:items_description}
\begin{small}
\begin{longtable}{@{\ts}lcp{7cm}c@{\ts}}
    \toprule
    Construct & Item ID & Questions & Reference \\ 
    \midrule
    Need for  & AN\_1 & (*) I was free to do things my own way. & \cite{SDT} \\ 
    autonomy  & AN\_2 & (*) I had a lot of pressures I could do without. & \cite{SDT} \\ 
              & AN\_3 & My choices expressed my ‘‘true self’’. & \cite{SDT} \\ 
              & AN\_4 & (*) There were people telling me what I had to do. & \cite{SDT} \\ 
              & AN\_5 & I was really doing what interests me. & \cite{SDT} \\ 
              & AN\_6 & (*) I had to do things against my will. & \cite{SDT} \\ 
    \hline
    Need for    & CN\_1 & I did well even at the hard things. & \cite{SDT} \\ 
    competence  & CN\_2 & (*) I experienced some kind of failure, or was unable to do well at something. & \cite{SDT} \\ 
                & CN\_3 & I took on and mastered hard challenges. & \cite{SDT} \\ 
                & CN\_4 & (*) I did something stupid, that made me feel incompetent. & \cite{SDT} \\ 
                & CN\_5 & I was successfully completing difficult tasks and projects. & \cite{SDT} \\ 
                & CN\_6 & (*) I struggled doing something I should be good at. & \cite{SDT} \\ 
    \hline
    Need for     & RN\_1 & I was lonely. & \cite{SDT} \\ 
    relatedness  & RN\_2 & I felt a sense of contact with people who care for me, and whom I care for. & \cite{SDT} \\ 
                 & RN\_3 & I felt close and connected with other people who are important to me. & \cite{SDT} \\ 
                 & RN\_4 & (*) I felt unappreciated by one or more important people. & \cite{SDT} \\ 
                 & RN\_5 & (*) I felt a strong sense of intimacy with the people I spent time with. & \cite{SDT} \\ 
                 & RN\_6 & (*) I had disagreements or conflicts with people I usually get along with. & \cite{SDT} \\ 
    \hline
    Home         & HWE\_1 & (*) My home is free from excessive outside noise. & \cite{wdq} \\
    working      & HWE\_2 & My job takes place in a clean environment. & \cite{wdq} \\
    environment  & HWE\_3 & (*) The climate at at my home is comfortable in terms of temperature and humidity. & \cite{wdq} \\
                 & HWE\_4 & The seating arrangements at my home are adequate (e.g.: ample opportunities to sit, comfortable chairs, good postural support). & \cite{wdq} \\
                 & HWE\_5 & My home offers an environment where I can easily focus on my work & \cite{office_satisfaction} \\
                 & HWE\_6 & There is enough space for work material and equipment at my home. & \cite{office_satisfaction} \\
                 & HWE\_7 & (*) At home, I have the technical equipment to do my work (e.g.: PC or laptop, printer, webcam, microphone). & \cite{pandemic_predictors} \\
                 & HWE\_8 & (*) The internet connectivity is fast and reliable at my home. & \cite{pandemic_predictors} \\
                 & HWE\_9 & (*) On the computer or laptop I use while working from home I do have the software and access rights I need. & \cite{pandemic_predictors} \\
    \hline    
    Use of Scrum & US\_1 & (*) The Product Backlog is always prioritized. & \cite{Scrum_guide} \\
                 & US\_2 & (*) The Scrum Team consists of Product Owner, Scrum Master and Developers (Development Team), composed of 3 to 10 people. & \cite{Scrum_guide} \\
                 & US\_3 & (*) Within my Scrum Team, there are no sub-teams or hierarchies. & \cite{Scrum_guide} \\
                 & US\_4 & (*) Only the people within the Scrum Team internally decide the work on the Sprint. & \cite{Scrum_guide} \\
                 & US\_5 & The Product Owner is the only person responsible of creating and ordering items in the Product Backlog while estimating the items is the Developers' responsibility. & \cite{Scrum_guide} \\
                 & US\_6 & (*) The Scrum Master is responsible to facilitate all the Scrum events. & \cite{Scrum_guide} \\
                 & US\_7 & (*) Sprints have fixed lengths that do not exceed one calendar month. & \cite{Scrum_guide} \\
                 & US\_8 & (*) Sprint planning is completed at the beginning of each Sprint and the resulting plan is created by the collaborative work of the entire Scrum Team. & \cite{Scrum_guide} \\
                 & US\_9 & (*) Daily Scrum is time-boxed to 15 minutes and it is held every day of the Sprint. & \cite{Scrum_guide} \\
                 & US\_10 & (*) During the Sprint Review, the Scrum team and the stakeholders collaborate on what was done in the Sprint and what to do next. & \cite{Scrum_guide} \\
                 & US\_11 & (*) The Sprint Retrospective occurs after the Sprint Review and prior to the next Sprint Planning and The Scrum Team discusses what went well during the Sprint, what problems it encountered, and how those problems were (or were not) solved. & \cite{Scrum_guide} \\
                 & US\_12 & Each artifact contains a commitment to ensure it provides information that enhances transparency and focus against which progress can be measured (Product Backlog - Product Goal, Sprint Backlog - Sprint Goal, Increment - Definition of Done). & \cite{Scrum_guide} \\
                 & US\_13 & (*) Only the Development Team can update the Sprint Backlog during the Sprint. & \cite{Scrum_guide} \\
                 & US\_14 & (*) Refinement of the Product Backlog items is done at least once each Sprint. & \cite{Scrum_guide} \\
                 & US\_15 & Each Increment is checked if it meets the Definition of Done before being deployed. & \cite{Scrum_guide} \\
    \hline
    Project success & PS\_1 & In light of new business requirements that arose during project execution, the project delivers all desirable features and functionality. & \cite{Agile_success_model} \\
                    & PS\_2 & In light of new business requirements that arose during project execution, the Scrum-developed software meets key project objectives and business needs. & \cite{Agile_success_model} \\
                    & PS\_3 & In light of new business requirements that arose during project execution, the Scrum-developed software overall is very successful. & \cite{Agile_success_model} \\
    \bottomrule \\
    \caption{Items description. Those prefixed with (*) were dropped because of their insufficient loading onto their latent variable}
    \label{table:items}
\end{longtable}
\end{small}

\section{Sample Study}\label{apx:sample_study}
\begin{figure}[h]
    \centering
    \includegraphics[width=\textwidth]{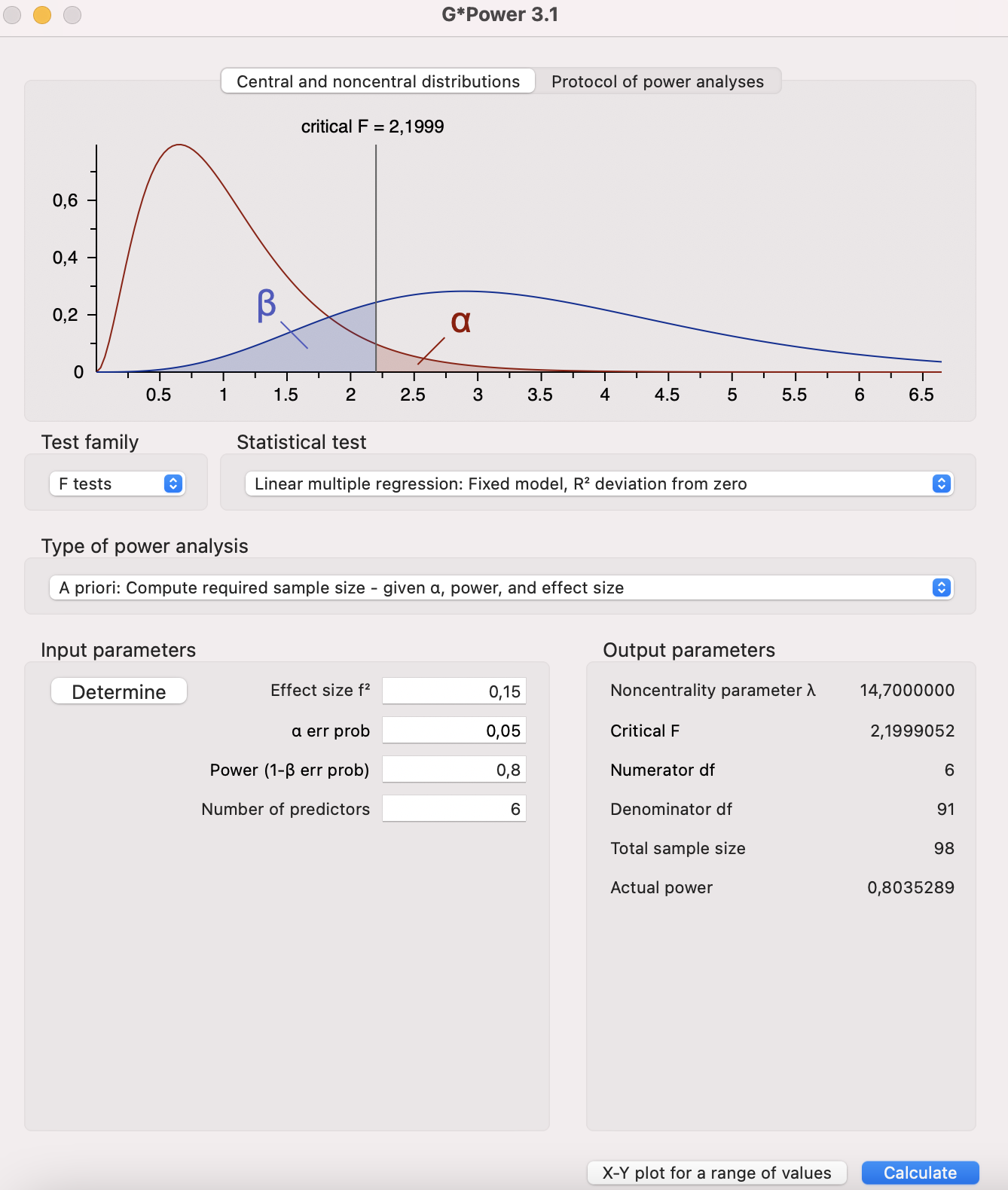}
    \caption{Sample size estimation based on statistical power of 80\%}
    \label{fig:g_power}
\end{figure}

\begin{table}[]
    \centering
    \small
    \begin{tabular}{@{\ts}l@{\ts[4.5]}c@{\ts[1.5]}c@{\ts[1]}}
    \toprule
               & Frequency & Percent (\%) \\ \midrule
    Man        & 112       & 81.2 \\
    Woman      & 25        & 18.1 \\
    Non-binary & 1         & 0.7  \\
    Total      & 138       & 100.0   \\ \bottomrule
    \end{tabular}
    \caption{Population divided per gender}
    \label{table:gender}
\end{table}

\begin{table}[]
    \centering
    \small
    \begin{tabular}{@{\ts}l@{\ts[2.5]}c@{\ts[1.5]}c@{\ts[1]}}
    \toprule
                                                         & Frequency & Percent (\%) \\ \midrule
    Portugal                                             & 37        & 26.8      \\
    United Kingdom                                       & 24        & 17.4      \\
    Poland                                               & 16        & 11.6      \\
    Italy                                                & 11        & 8.0       \\
    Spain                                                & 7         & 5.1       \\
    Mexico                                               & 6         & 4.3       \\
    Canada                                               & 6         & 4.3       \\
    Greece                                               & 4         & 2.9       \\
    South Africa                                         & 4         & 2.9       \\
    Chile                                                & 3         & 2.2       \\
    Austria                                              & 2         & 1.4       \\
    Czech Republic                                       & 2         & 1.4       \\
    Denmark                                              & 2         & 1.4       \\
    France                                               & 2         & 1.4       \\
    Germany                                              & 2         & 1.4       \\
    Norway                                               & 2         & 1.4       \\
    United States of America                             & 2         & 1.4       \\
    Slovenia                                             & 2         & 1.4       \\
    Belgium                                              & 1         & 0.7       \\
    Hungary                                              & 1         & 0.7       \\
    Latvia                                               & 1         & 0.7       \\
    Netherlands                                          & 1         & 0.7       \\
    Total                                                & 138       & 100.0      \\ \bottomrule
    \end{tabular}
    \caption{Country of residence}
    \label{table:country}
\end{table}

\begin{table}[]
    \centering
    \small
    \begin{tabular}{@{\ts}l@{\ts[1.6]}c@{\ts[1.5]}c@{\ts[1]}}
    \toprule
                                 & Frequency & Percent (\%) \\ \midrule
    Developer (Development Team) & 119       & 86.2 \\
    Product Owner                & 7        & 5.1  \\
    Scrum Master                 & 6        & 4.3  \\
    Other                        & 6        & 4.3 \\
    Total                        & 138       & 100.0   \\ \bottomrule
    \end{tabular}
    \caption{Scrum role}
    \label{table:scrum_role}
\end{table}

\begin{table}[]
    \centering
    \small
    \begin{tabular}{@{\ts}l@{\ts[3.35]}c@{\ts[1.5]}c@{\ts[1]}}
    \toprule
                       & Frequency & Percent (\%) \\ \midrule
    Less than 1 year   & 39        & 28.3        \\
    1-2 years          & 40        & 29.0        \\
    3-5 years          & 43        & 31.2        \\
    6-10 years         & 15        & 10.9         \\
    11-20 years        & 1         & 0.7         \\
    More than 21 years & 0         & 0.0         \\
    Total              & 138       & 100.0          \\ \bottomrule
    \end{tabular}
    \caption{Scrum experience in years}
    \label{table:scrum_exp}
\end{table}

\begin{table}[]
    \centering
    \small
    \begin{tabular}{@{\ts}l@{\ts[1.55]}c@{\ts[1.5]}c@{\ts[1]}}
    \toprule
                                 & Frequency & Percent (\%) \\ \midrule
    During the COVID-19 lockdown & 100       & 72.5       \\
    Before the COVID-19 lockdown & 38        & 27.5       \\
    Total                        & 138       & 100.0      \\ \bottomrule
    \end{tabular}
    \caption{Start working from home}
    \label{table:start_wfh}
    \end{table}

\end{document}